\documentclass[12pt]{book} 

\usepackage{cranfieldthesis}
\usepackage{graphicx}
\usepackage{multirow}
\usepackage[utf8]{inputenc}
\usepackage{amsmath}
\usepackage{amssymb}
\usepackage{listings}
\usepackage{color}
\usepackage{float}

\title{EU H2020 Gauss project. Geo-Fencing Software System}
\author{Hao Xu}
\date{August 2018}
\school{Aerospace, Transport and Manufacturing}
\course{Aerospace Vehicle Design }
\degree{MSc}
\academicyear{2017--2018}
\supervisor{Dr Huamin Jia}
\copyrightyear{2018}

\begin{document}

\frontmatter

\maketitle

\begin{abstract}
	
The Geofencing system is the key to operate the Unmanned Aerial Vehicle (UAV) within the safe and appropriate zone to avoid public concerns and other privacy issues. The system is designed to keep the UAV away from geofenced obstacles using the onboard GNSS and IMU location.
The Geofencing system is part of the H2020 GAUSS project and facilities other subsystems, for instance, to support the command and control link, which is the security measure to secure the UAV from hijacking and signal spoofing. 

The regulatory authorities expressed the concern of having UAVs flying in the no-fly zone and causing troubles from offending private privacy to hazards at airport airspace. Hence the geofence system shall provide guidance message, which enables the UAV to evacuate from no-fly-zone, based on real-time updated location.

This thesis aims to first illustrate the generation of geofence and then apply the geofence system on UAV operation. This application enables UAV to fly in the designated area without human intervention. The project is built with JAVA using GIS-enabled Database Management System and Open Soured Map data powered by OpenStreetMap and OS map. This method has been tested by simulations which had results of high accuracy.

    \section*{Keywords}
    Geo-fence; 3D Map; Autonomous Vehicle; OpenStreetMap; PostGIS; GDAL;
\end{abstract}

\sstableofcontents

\sslistoffigures

\sslistoftables

\begin{listofabbreviations}
\abbrev{API}{Application Programming Interface}
\abbrev{BLOB}{Binary Large Object}
\abbrev{C2C}{Command and Control}
\abbrev{DBMS}{Database Management System}
\abbrev{DOM}{Document Object Model}
\abbrev{GDAL}{Geospatial Data Abstraction Library}
\abbrev{GEOS}{Geometry Engine, Open Source}
\abbrev{GIS}{Geographic Information System}
\abbrev{GNSS}{global navigation satellite system}
\abbrev{GPL}{GNU General Public License}
\abbrev{GUI}{Graphical User Interface}
\abbrev{JDBC}{Java Database Connectivity}
\abbrev{LBS)}{Location Based Service}
\abbrev{MAV}{Micro Aerial Vehicle}
\abbrev{MTOW}{Maxium Take Off Weight}
\abbrev{OSG36}{the national grid system}
\abbrev{OSM}{OpenStreetMap}
\abbrev{PNG}{Portable Network Graphics}
\abbrev{POSIX}{Portable Operating System Interface}
\abbrev{SATM}{School of Aerospace, Technology and Manufacturing}
\abbrev{SRID}{Spatial Reference System Identifier}
\abbrev{UAV}{Unmaned Aerial Vehicle}
\abbrev{UAS}{Unmanned Aerial System}
\abbrev{USGS}{US Geological Survey}
\abbrev{WKB}{Well-Known-Binary}
\abbrev{WKT}{Well-Known-Text}
\abbrev{XML}{Extensible Markup Language}
\abbrev{}{}
    
\end{listofabbreviations}

\chapter{Acknowledgements}
I would like to thank my supervisor Dr. H. Jia, for his kind support across the project. I appreciate the effort in making me develop a comprehensive understanding of software development and geographic information system. 

My thanks to all my colleagues and friends who have provided their kind support on this project. 

I would like to thank my parents who made my study at Cranfield possible.

%
%
\mainmatter

\chapter{Introduction\label{cp:1}}
\section{Background}
Geofencing is described as a virtual definition of a real-world geographical space, including its perimeter and height. It is artificially generated based on users or regulators interests. It combines awareness of the object's current location with awareness of the object's close to the point of interest. 

\begin{figure}[H]
\centering
\includegraphics[scale=0.4]{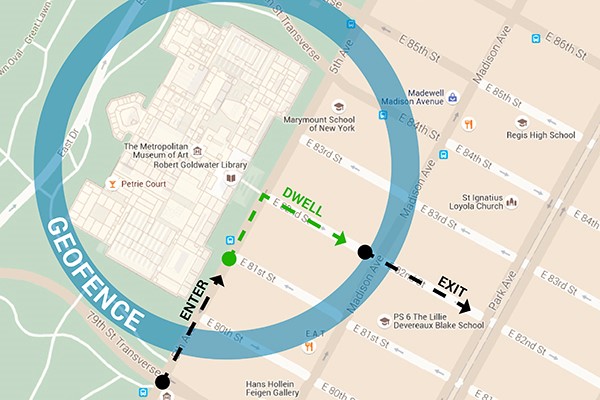} 
\caption{The Geo-Fence Concept}
\label{fig:1}
\end{figure}

As shown in Figure \ref{fig:1}, the moving object is watched by the geofence system, where the blue circle lays. The circle acts like a fence logically, and can be triggered based on users preferences. 
For example, once the target hits the boundary defined in the GIS, a response is geo-triggered. A series of motions can be taken into account based on the event type, including $``$Inside", $``$Outside", $``$Entry", $``$Exit" and $``$Proximity".

\section{Motivation}
H2020 GAUSS project has a wide scope of achievable targets. The Geofence is developed as the supporting point of the Command and Control (C2C) communication system.

\begin{figure}[H]
	\centering
	\includegraphics[scale=0.5]{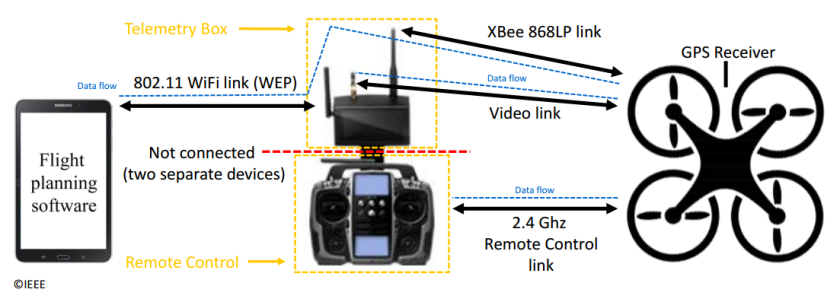}
	\caption{Data link for Command\&Control of a typical UAV}
	\label{fig:2}
\end{figure}

Command and Control link encryption for UAV represent the uplink and downlink encryption of UAV commands. It was revealed in the past how fragile the link is. In 2011, the incident of RQ-170 US drone hijack revealed the vulnerability of the UAV guidance system and the datalink used by the US military. The encryption of the datalink shall prevent the spoofing and deception from the unauthorised sources. 
There are two types of links illustrated in Figure \ref{fig:2}, the Command and Control link is generally an analogue signal without encryption but a spread spectrum with a certain sequence of channels, however, the message itself was not encrypted. 

The increasing UAV spoofing event has alerted the authority especially military and law enforcement agencies, where the techniques they used in past has been turned against them. The need for an end-to-end encryption is demanding trend of UAV Comm-link. Nevertheless, the extent of this project has been restrained to geofencing rather than the discussion of the data-link, hence the data-link is not part of the work but assumed as the designated environment.

The Communication flaws lead to the requirement of UAV self-restrains, in a word, geofence. The geofence helps the UAV stay in safe and cleared airspace in the event of signal breach or lost. This forms the motivation of the Geofence Software development.

\section{Objectives}
This thesis focus on solving the geofence software that can be used by the UAV while flying cross certain areas and objects.

The Geofence is designed for the UAV flying with above equipment, therefore, it not only produces geofences but also guidance messages. The directing message is generated using the situational information from the database.

The feature of this project is its stand-alone architecture, which does not require the internet connection and works on the sole machine.

Besides, the software is developed and deployed with Open Sourced Tools and Libraries that do not require special agreements to run and distribute. 

The objectives for the geofence development, include:

\begin{itemize}
	\item Geofence Advisory
	\item Geofence Situation
	\item Geofence Display
	\item Geofence Architecture
	\item Geofence Interface
	\item Detailed design of geofence functions
	\item Performance test result of the system
	\item Simulation with the realistic condition
\end{itemize}

\section{Chapter Organisation}
\textbf{Chapter \ref{cp:1}} is the introduction of the project, it explains the motivations and the intended functions of Geofence \newline
\textbf{Chapter \ref{cp:2}} introduces the discovery and necessary background towards the Geofence Software. It covers Geofencing concept, Advisory logic, Geographic Information System, Map reference, open-sourced Map provider and UAV types that are relevant to the Geofence Software. It also briefly covers the Map render background and a sample of making a Graphical User Interface. The knowledge stated in Chapter 2 is necessary for people who have no experience of software engineering, Geographic Information System, and UAV sector of Aviation Industry.\newline
\textbf{Chapter \ref{cp:3}} explains the methodology used in the project development, highlighting software development pattern. Beyond the project development, it also features the philosophy of the development, the ethos of a modern programmer, together with rational of making choice on language, database and beyond.\newline
\textbf{Chapter \ref{cp:arch}} shows the architecture and the design of the whole system. It explains the reason why the architecture is suitable for the Geofence and also compares it with other potential architecture.\newline
\textbf{Chapter \ref{cp:5dev}} is the main part of the thesis, it analyses the development stages and classifies the whole software into 3 parts, the Interface, the Geofence and the Database. In the end, a brief introduction on software compiling is stated but not going deep into it.\newline
\textbf{Chapter \ref{cp:6test}} covers the test and set up of the Geofence Software. It also analyses the performance issue of the Geofence Software, which is very important for real-time operation.\newline
\textbf{Chapter \ref{cp:7con}} summaries up the thesis, provides the comment of the project and also the limitations. In addition, it states the future work to make this project more competent among others.

\chapter{Literature Review\label{cp:2}}
A complete Geofence system requires works from many disciplines of studies, including Geographic Information System, Database Management System, Geometry Study, Cartography, Topology, UAV flying modes and Computer graphics for map rendering.
\section{Scope}
The study of Geofence is in fact the study of knowledge collection mentioned above and assembles them into a functioning program that enables geofencing to work in the environment that the user desires.

In the following pages, the tools and knowledges related to Geofencing will be reviewed.
\section{Geofencing}
\subsection{Rule of Geofencing}
The Geofence is usually constructed by a set of rules, including the limiting area defined by coordinates of points of interest and the entry/exit rules of the designated area. The rules are mostly defined by the regulatory body and the operator/land-owners' will. 

The commercial Location Based Service(LBS) is the most principle use of such technology, however, not in the greater extent. The use of geofence can also be found on recently grossing business models of sharing economy, for example, sharing bicycles and sharing cars. These service providers demand a well-functioned geofence system to keep their assets in the wanted area, instead of being relocated to the unfavoured area or location. A common usage of Geofence for bicycle sharing can be found in bicycle parking, where the location is vital for the company and municipal government. If the bicycle is intended to be left on the unwanted area, the lock will stop users to park the bicycle and even penalise the user for such behaviours.   

\begin{figure}[H]
	\centering
	\includegraphics[scale=0.3]{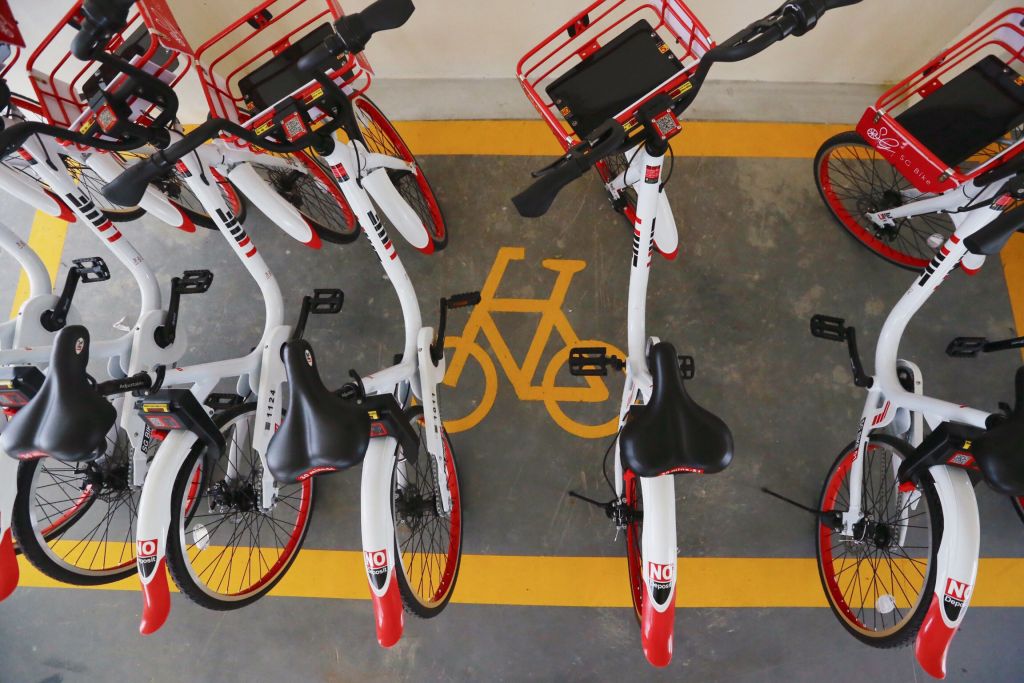}
	\caption{A geofence scenario of sharing bike parking\cite{EDWARDLIMYEWSIH2017}}
\end{figure}

GAUSS project defines the Geofence as a synthetic concept, using the parameters like building height, land use, designated area, to form a Geofenced Flying Zone. The building type attribute can be used as the partition parameter, for example, the type hospital will be classified as obstacles, but the type warehouse is accessible under normal settings.

The Geofence will create a safer place for people on the ground and also lead to a better operating condition.

\subsection{Collision control}
The collision damage caused by any flying object is a great concern for all circumstance. UAVs, made by anyone, have much higher chances of losing control and incurring safety issues. The civil air regulation does cover the field of UAVs and Micro Aerial Vehicle (MAVs). The damage of a falling object various from the object's mass, however, even the lightest one could hit the ground personnel with a considerable high speed and cause injury. Hence, the collision control of the UAVs should be reinforced by flying rules, restrictions on regions and Geofence.

The Geofence can help the UAV fly in the safe and open region, without putting buildings, facilities and people in danger. By setting up a Geofence, the denial of entry stops the UAV flying above the certain items, for example, residential buildings, schools and playgrounds for children. The Geofence can divert the vehicle from entering it and also indicates the path to an open area. 
\subsection{Geofence Advisory}
The geofence works with the advisory mechanism, where the diverting course and Estimated Time of Arrival (ETA) will be produced for the end users' reference. 

The UAV itself is a moving object while flying along a course, without changing headings and speed, the ETA to the verge of next object is solvable. Before the UAV enters or collide with it, the advisory shall provide a solution to avoid it.

The solution may contain advised velocity and divert course for the UAV or simply give a Stop instruction. Due to the inertia of the body, the UAV needs a buffer zone to stop or manoeuvre, in order to keep a safe distance from the no-fly zone. 

The advisory information can be audio, visual or text based. The UAV operator shall choose the way how it is presented. The text based message shall never appear in the UAV operation, due to safety concern, and the visual alert shall correspond avionics colour conventions. For most situation, audio alert is preferred among others. Selective attention makes sure the human ears only hear one sound at a time, even multiple sounds are sensed\cite{Novacek2003}. 
\subsection{Height from multiple sources}
To fly a UAV, the height information is the most important factor, in terms of safety and regulation.
The airspace is regulated by the height and it is essential to fly under regulations. The Geofence is also partly defined by the height, where the operators and specific lands require new height limits for the region. 

The height comes from many sources and can be selected from one source at a time. The different height sources may disagree with each other and cause misunderstanding or fault.
From the GNSS, an ellipsoid height, which is not really the height relevant to any object but a conceptual height based on the earth model, is obtained. The terrain height from the land survey uses different models to fix the height from GNSS. UAV flies with embedded barometer and radio altimeter, in order to obtain height in its own favour. In the meantime, the building height is supplied in the form of height above ground, which is very different from all the heights mentioned above.
\subsubsection{Geoid Height}
\begin{figure}[H]
	\centering
	\includegraphics[scale=0.71]{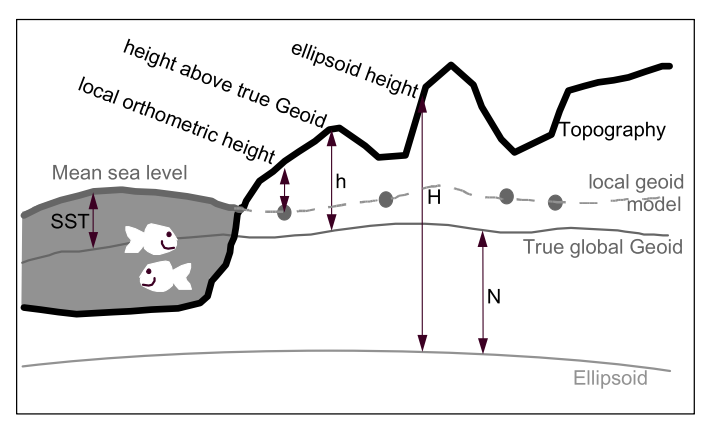}
	\caption{The relation of Geoid and local Geoid model compare with mean sea level and ellipsoid panel\cite{OrdnanceSurvey2018}}
	\label{fig:geoid}
\end{figure}

The term Geoid height is the distance above a reference panel, which compensate the effect of ellipsoid reference. In the ellipsoid reference, as shown in Figure \ref{fig:geoid}, the surface is not flat, hence the length above it does not mean the height. The conversion between Geoid height and ellipsoid height is shown as Equation \ref{equ:height}:
\begin{equation}
H = h + N
\label{equ:height}
\end{equation}
where N is the "Geoid-ellipsoid separation".\cite{OrdnanceSurvey2018}  The N, defined by the Geoid Model, varies from places to places. The development of Geoid allows the GNSS to provide more accurate height information since the output height from GNSS is based on Ellipsoid model.

The Geoid model is designed using the local terrain information, where the level should correspond to the sea level. The height will be slightly different due to the inconstant gravity changes. Therefore, many Geoid Models are used by different countries and regions to improve the local height accuracy.

The Geoid hence solved the height problem in the sense of geography. In this case, the unified height can be introduced with confidence.
\subsection{Existing Geofence Solution for UAV}
There are several UAV Geofence work published recently in Pratyusha's \cite{Pratyusha2015} work. Praatyusha's paper described a set of algorithms to work with UAV Geofence. The way he managed is to set up polygonal shapes as the geofence and detect UAV entry \& exit events. His thought is illuminative and works he has done upon it has provided approaches to handle UAV against geometrical shapes. However, his study is limited to the UAV handling in a manually set up area, with manual geofence generation. It has limited usage against real situations.

\section{Geographic Information System}
Geographic Information System(GIS) is the terminology to process and manage geographical spatial data. It is a subject under geography science and now has been used for many fields including but not limited to map industry, but also agriculture, game making and outdoor sports.
\begin{figure}[H]
	\centering
	\includegraphics[scale=0.31]{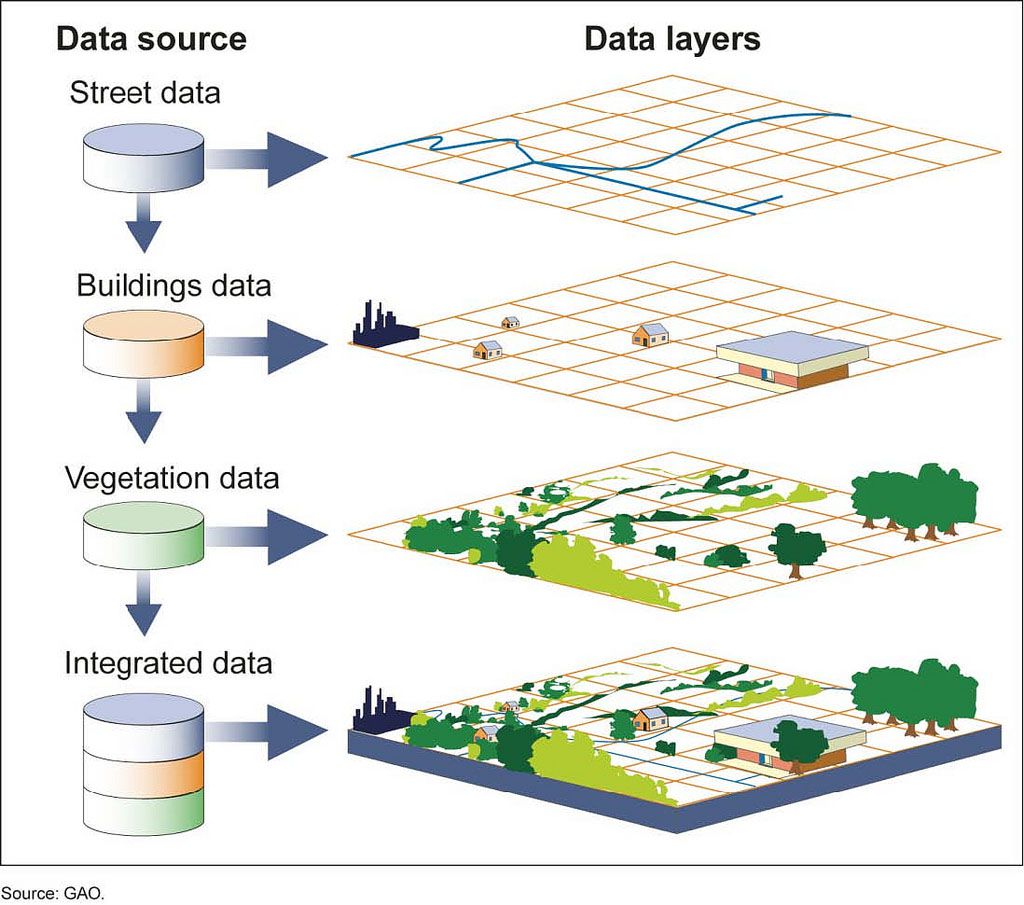}
	\caption{Layers of GIS\cite{U.S.GovernmentAccountabilityOffice2018}}
\end{figure}

\subsection{Spatial Boundary Generation and Slicing}
Well-Known-Binary (WKB) and Well-Known-Text (WKT) are the two most common ways to represent a geo-related geometry information, as indicated by their names.   

The geometry is generated either from a defined point with parameters that tell the shape, radius or perimeters, or a polygon shape defined with multiple longitudes, latitude rally points.  

The following picture and polygon are generated using WKB geometry:
\begin{figure}[H]
	\centering
	\includegraphics[scale=1]{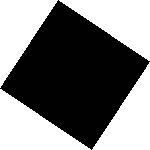}
	\caption{A 150*150 px PNG of a WKB geometry}
\end{figure}

The WKT message retrieved from above geometry reads as:MULTIPOLYGON(((13.7244306 51.0336413,13.7245794 51.033782,13.7248143 51.0336837,13.7246655 51.033543,13.7244306 51.0336413))) 

In this text information, four points are defined as the regulating points of this shape. It contains the coordinates which were agreed prior to the data extract.
\subsection{Map in the digital world}
Cartography is the science of making maps, however, people draw it differently.

The major difference is the reference system used for the unit and the projection. The modern reference systems are based on two principles, the grid and the long-lat system.

The map sources involved in this study are either using WGS-84 or British National Grid Reference. WGS-84 is commonly used for GPS application and international Map making, the later one, is only used in Britain and has a little difference. 

The grid puts the UK map in many boxes, which are 500km $\times$500km\cite{OrdnanceSurvey2018} on the scale of the first letter, each box is then divided into smaller ones by 100km $\times$100km, named with two letters. The reference uses the number to indicate which smaller box is under the larger one. Additionally, the number of digits shows how detail the location is described.
\begin{figure}[H]
	\centering
	\includegraphics[scale=1]{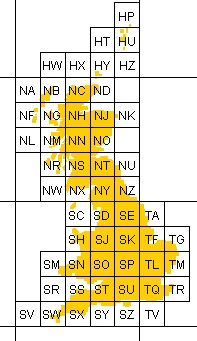}
	\caption{The Grid of the UK\cite{OrdnanceSurvey2018}}
\end{figure}
The number comes after the first two letters, with range from 0 to 10, on the horizontal axis, then the second half represents the vertical axis. As mentioned before, each two-letter box is 100km $\times$ 100km, which brings the subsequential box 10km long and wide resulting 10km $\times$ 10km resolution.

\begin{figure}[H]
	\centering
	\includegraphics[scale=1]{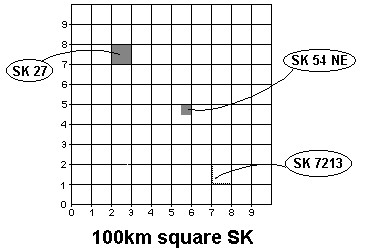}
	\caption{Example of SK grid\cite{OrdnanceSurvey2018}}
\end{figure}
An example of a six-digits grid reference, SK123456, means in the SK block, 123km easting and 456km northing from the origin of grid SK, and grid SK is 2500 km east and 2000km north from the origin of the system. From six digits reference, the resolution is improved to 1km.

The National Grid Reference is standardised as OSGB36, which was introduced in 1936. A geodetic transformation between OSGB36 and other systems is possible, whereas, the typical error between OSGB36 and WGS84 can range from 7m to 120m depending on the method attempted and the area converted.

A commonly used framework of the spatial reference system, ESPG, introduces a range of codes to manage them, called ESPG code. The OSGB36 is ESPG:27700 and WGS84 is ESPG:4326.

These codes are used in the Spatial Reference System Identifier(SRID) framework, and each of them is stored in the spatial database along with the geometric data.
\subsection{Open Sourced Map data and terrain data}
\subsubsection{DEM-Global Elevation Data}
Elevation data supply height above the sea level for every tile in the resolution of 30x30 meters. The raw data from the US Geological Survey(USGS) and European Union DEM have been adapted to fill the map using a similar approach as other open map standards.

\begin{figure}[H]
	\centering
	\includegraphics[scale=0.7]{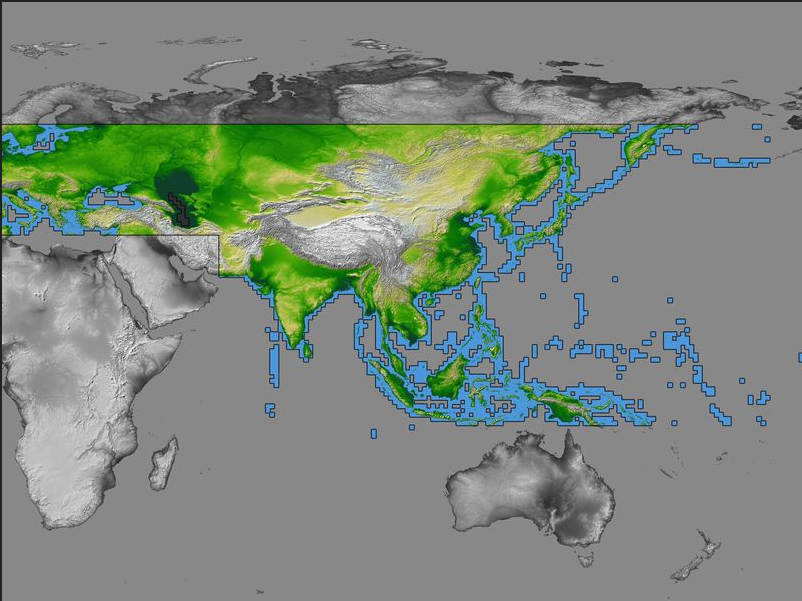}
	\caption{Sample of Euroasia SRTM mission\cite{JetPropulsionLaboratory2004}}
	\label{fig:srtm}
\end{figure}

USGS Shuttle Radar Topography Mission (SRTM) is the public domain source of earth elevation obtained from STS-99 Shuttle Mission. A sample image of Euroasia area is shown in Figure \ref{fig:srtm}

The model was established from tiles based on the WGS84 reference, each of them has one degree of longitude and latitude. The resolution from the raw data is 1 arc-second, which is restricted for government uses. The publicly accessible data is 3 arc-second.

\subsubsection{OpenStreetMap}
OpenStreetMap(OSM) is a map of the world, created and curated by people collectively. It is free to use and free to use and edit. The data generated by the project is the main object, and the creation of such content has happened around the world. Since the data is managed completely by the people, the spread of the map has gained strong positive feedback from the users as a great alternative to the closed-source map suppliers, for instance, google map. 

\begin{figure}[H]
	\centering
	\includegraphics[scale=0.6]{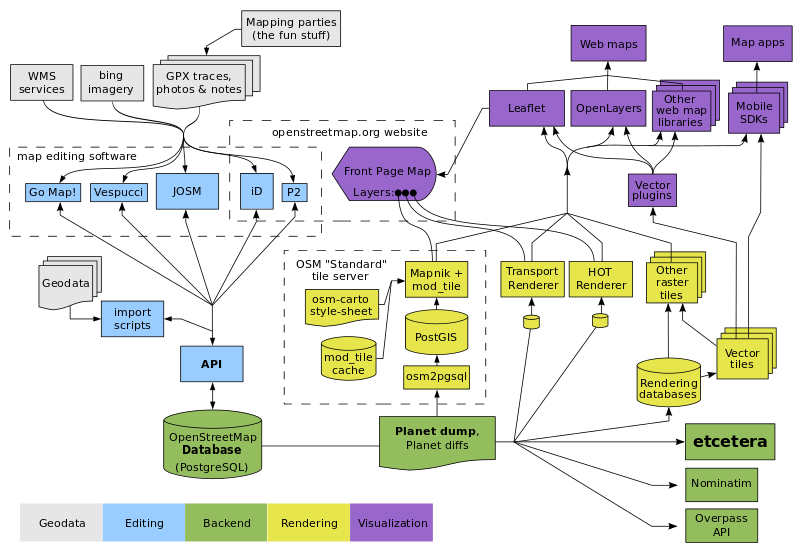} 
	\caption{The components of OpenStreetMap\cite{OpenStreetMapContributors2017}}
	\label{fig:osm}
\end{figure}

The development of the OSM project has included not only the content but also the product chain from map creation tools and several useful map tools for map rendering and data management.  In Figure \ref{fig:osm}, an overall architecture of the OSM project is colour coded into five functions: Data, Editing, Website Backend, Map rendering and the Visualisation. It is a great example combination of a huge set of open-sourced tools, and forms the great tool chain for GIS work and application.

OSM map heavily relies on the public domain map resources including the British Ordnance Survey Map, USA Census TIGER data of the States. Beyond the public domain contribution, the commercial map supplier has also lent a helping hand to allow the topography map to be matched with Satellite Imagery. Moreover, personal knowledge has been one of the largest sources of input, such as, the number of houses, the functionality of the buildings, opening time information, etc. The map truly is a collective effort and will be improved by its enthusiastic community.   

It has been used in academic and low-cost commercial solution, for its advantages of being open-sourced and free to use. 

The map is now the backbone of several essential Location Based Service(LBS) applications. On the top of that, the route planning tools are also developed and licensed as open-source.

\subsubsection{Mbtiles from OpenMapTiles}
The Mbtiles, which is generated by OpenMapTiles organization, has used the data imported from OpenStreetMap.  It converts the XML format data into a binary form, tile by tile. It stores all binary data in the SQLite database, hence improves the speed of indexing and accessing.
\subsection{Unit Convention}

In a Latitude-Longitude system, degree is the unit for the coordinates, however, computers can only read digits.

Hence the representation of the coordinate is in the form of a float number. Every digit after the dot are float numbers instead of minutes and seconds. 

The unit conversion of the degree to digits is done using the simple equation below:

\begin{equation}
	1 degree =60 minute=3600 seconds
\end{equation}
0.5 degree is 30 minutes, for example, 50.5 degree is 50 degree for 30 minutes. 

Sexagesimal is then converted into decimal.

To covert the unit from a sphere-based system to a Cartesian system, it requires geodetic transformation, which causes an error in the converted system. 

Helmert datum translation \cite{coordinateguide2018} explains a method to work out the difference of two perfect mathematical reference system. In practice, the error of the model is fixed by observing known real ground points, and hence fixes the error in the certain area. In another word, the fix can only apply to a rather small area. 

The coordinate transformation is performed by equation \ref{equ:coordinate}:
\begin{equation}
\label{equ:coordinate}
{\left[\begin{array}{c}x\\ y\\ z\end{array}\right]}^{B}
={\left[\begin{array}{c}t_{x}\\ t_{y}\\ t_{z}\end{array}\right]}
+\begin{bmatrix}
1+s & -r_{z} & r_{y} \\ 
r_{z} & 1+s & -r_{x} \\ 
-r_{y} & r_{x} & 1+s
\end{bmatrix} \ast {\left[\begin{array}{c}x\\ y\\ z\end{array}\right]}^{A}
\end{equation}
where $t_{x}$, $t_{y}$, $t_{z}$ are the translation along the X, Y, Z axis in meters and $r_{x}$, $r_{y}$, $r_{z}$ are the rotation about the X, Y, Z axis in radians and $s$ is the scale factor minus one\cite{OrdnanceSurvey2018}, which is often stated in parts per million.

\begin{figure}[H]
	\includegraphics[scale=0.7]{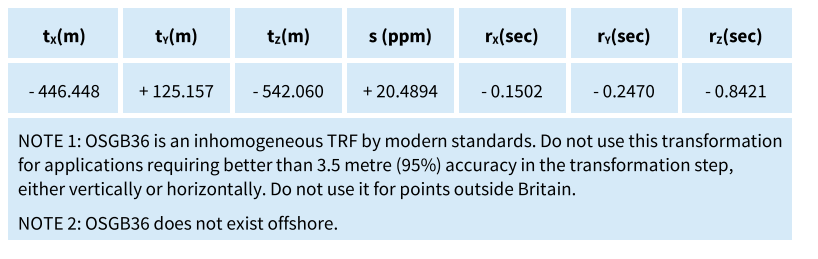}
	\caption{Approximate WGS84 to OSG36 transformation\cite{OrdnanceSurvey2018}}
	\label{fig:osg}
\end{figure}

\begin{figure}[H]
	\includegraphics[scale=1]{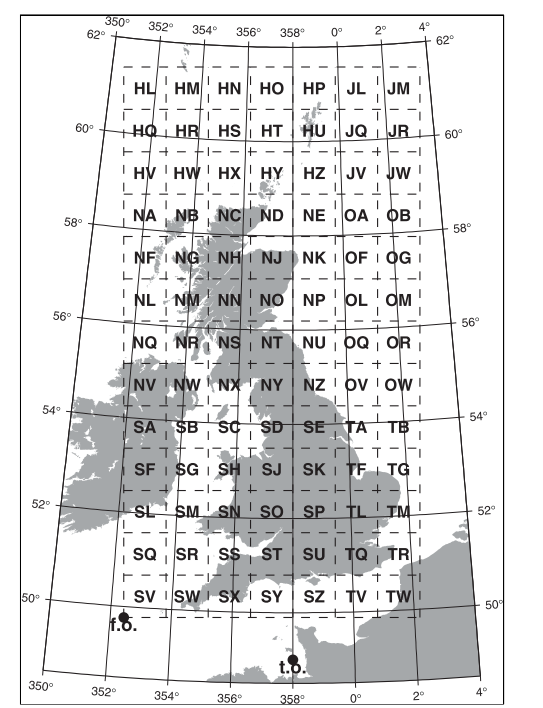}
	\caption{The National Grid, showing the true origin (t.o.) and false origin (f.o.)\cite{OrdnanceSurvey2018}}
	\label{fig:grid}
\end{figure}

In Figure \ref{fig:osg}, a set of parameters is intended to give the Helmert transformation in approximately 3.5 meters error in 95 \% of the defined area of OSG36. In Figure \ref{fig:grid}, a Cartesian system, OSG36, and a spherical project system, WGS84, are compared.

There are ready to use algorithms available in the tools used in the project development. The important thing is the correct Terrestrial Reference System must be used in the appropriate operation.

The operation of the WGS84 will return the result in a degree and the transformation of the WGS84 to a meter-based system will return results in meters.

Transformation of WGS84 to OSG36 is the key for the UK geofence application, however, osg36 only covers the land and sea near the UK. For the use in America and other places in the world, EPSG:3857, WGS84 Pseudo-Mercator is required. Additionally, there are many sub area terrain reference systems, for instance, EPSG:26986, a model only covers Massachusetts, U.S.A. and uses NAD83 datum instead of WGS84. The main reason to use the sub level reference is because of higher accuracy, as mentioned earlier, the real ground points improve the accuracy dramatically. In Figure \ref{fig:grid}, a comparison of the national grid system(OSG36) and WGS84 are presented. The false origin, which is for the sake of convenience, is set at 400km west and 100km north of the true origin to ensure all the points in Britain are positive.\cite{OrdnanceSurvey2018}

\section{Unmanned Aerial Vehicle types}
UAVs are a general definition of the flying platforms without Man on-abroad. However, there are many criteria to class them into different categories. The terminology of UAV classification is out of the scope of this thesis, therefore an external reference is required. According to the U.S military Unmanned Aerial System (UAS) classification, there are 5 groups, separated by the weight, operating altitude and speed of the system. This criterion will be used through the project because the geofence is, firstly, height related and secondly, speed related.

\begin{table}[H]
	\centering
	\caption{UAS Classification\cite{UnitedStateDepartmentofDefense2011}}
	\label{tab:01}
	\begin{tabular}{|l|l|l|l|}
		\hline
		UAS Group & Maximum Take off Weight(kg)                                                                        & operating altitude (ft)                               & Speed (kn)                                                                 \\ \hline
		Group 1   & 0 - 20                                                                            & \textless 1,200                                       & 100                                                                        \\ \hline
		Group 2   & 21 - 55                                                                           & \textless 3,500                                       & \multirow{2}{*}{\begin{tabular}[]{@{}l@{}}\textless   250\end{tabular}} \\ \cline{1-3}
		Group 3   & \textless 1,320                                                                 & {\multirow{2}{*}{\textless 18000}} &                                                                            \\ \cline{1-2} \cline{4-4} 
		Group 4   & \multirow{2}{*}{\begin{tabular}[]{@{}l@{}}\textgreater   1,320\end{tabular}} & \multicolumn{1}{c|}{}                                 & \multicolumn{1}{c|}{\multirow{2}{*}{Any airspeed}}                         \\ \cline{1-1} \cline{3-3}
		Group 5   &                                                                                 & \textgreater{}18000                                   & \multicolumn{1}{c|}{}                                                      \\ \hline
	\end{tabular}
\end{table}

Table \ref{tab:01} shows the UAV categories, provided by DoD of States. The use of this table is not for authority reason but convenience. FAA regulates all the UAVs using Part 107, which regulates the operation rules and weight of the UAV. 

The introduction of the UAV classes is meant to create suitable manoeuvre decision. The main objects for discussion are the Group 1 and Group 2 UAV class since the geofence is not commonly considered where the operating altitude exceeds buildings height too much.

FAA generally requires the UAV to be weight less than 25kg and maximum ground speed of 100 km per hour, which falls into Group 1 and a part of Group 2.

\section{Spatial database}
Spatial Database organises the spatial object, which is defined in a geometric sense, in a more efficient way. The known example of using spatial database is the PostGIS integration with PostgreSQL database. It uses PostgreSQL as the engine to run the spatial extension, which deals with spatial type data. It enables the database to query the spatial data, for example, a polygon. It is not possible to index it in a standard database, but with the help of a spatial extension.

The spatial data are described in numerous forms, whereas, the raster and the vector data are the most common options among others.

The common raster database extension example is SQLite in the example of MBtiles.

The Vector database can be seen on the PostGIS on the top of PostgreSQL.

\subsection{GDAL and its methods}
The algorithms behind the extension are provided by the Geospatial Data Abstraction Library (GDAL)\cite{GDAL/OGRcontributors2018}, which is an open source project dedicated to the raster and vector geospatial data processing. The library is developed using C and C++ language. Its work has been used in many commercial products and other open source projects. 

It uses "Geometry Engine, Open Source" (GEOS) library and Proj4 library to support the spatial data manipulation, which is a robust and powerful tool. The Proj4 is an open source library of projection conversion. It helps the Geography reference system stays in line with each other and converts coordinates for external usages.  

Both libraries are required to be compiled with the host system to enable the GDAL functions. And the GDAL library needs to comply before using it in PostgreSQL and other software.

\begin{figure}[H]
	\centering
	\includegraphics[scale=0.9]{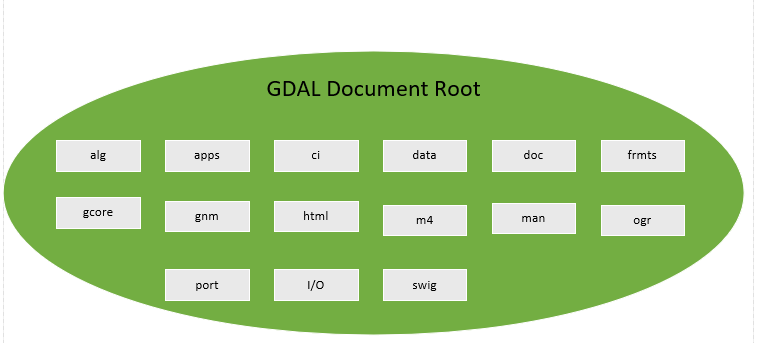}
	\caption{The file structure of GDAL Library source}
	\label{fig:file}
	
\end{figure}

Under GDAL source file structure, version 2.3.1, as shown in Figure \ref{fig:file}
\subsubsection{Brief Introduction of Folders and files:}

 alg: GDAL algorithms are stored in this folder\newline
apps: Command-line-Tool Source code collection\newline
bridge: Definition and links about GDAL abstract class and GDAL struct data.\newline
data: Configuration files are stored under data folder\newline
doc: Guidance for how to generate GDAL help document\newline
frmts: GDAL parser source code for different image formats.\newline
gcore: GDAL interface code for abstract classes, including data set, images, etc.\newline
html: The help documents generated by GDAL\newline
m4:  Not usable under windows development environment\newline
man: Help documents for the Unix-like system, not usable under windows development environment\newline
ogr: OGR library source code\newline
port: CPL library source code, for supporting the GDAL in the lower level. It defines the I/O and interfaces to another interface. \newline
swig: The swig script is used to pack the C/C++ library into Python, Java, and other language interfaces.
\\\\

To use GDAL, the library has to be built and compiled using build tools like Microsoft Visual Studio or CMake in the Linux environment.
\subsection{PostgreSQL database management system}
The PostgreSQL is introduced as the main database management system for this project, especially for the the spatial data processing. It is an object-relational database management system, originally developed by University of California. It plays an important role in GIS industry for its advanced features and history.
\subsubsection{Binary Large Object}
A Binary Large Object(BLOB) is a common form for images to be stored in most modern databases. The BLOB can be used for any undefined objects including audio, multimedia, etc. 
\subsubsection{POSTGIS}
PostGIS is a spatial database extender for PostgreSQL, which is the de facto development tool for spatial related software development by individuals and communities. It is commonly found in all Location-based-service service. Using the SQL language, it can easily manipulate spatial, geographical data.  It introduces geometry data type into the PostgreSQL database, facilitates viewing and editing GIS data.  The PostgreSQL integration allows web-based access for all GIS data, instead of desktop based tools.

\subsection{SQLite database management system with MBTiles Data}
SQLite is an in-process library bonded with a self-contained SQL database engine. It is the most common embedded database option and works well with most common programming languages and frameworks. It is selected as the database solution for MBTiles data, which is the default data format of OpenMapTiles service. 

MbTiles is a specification for storing the arbitrary tiled map in SQLite databases \cite{Fischer2018}. The data sheet is built with SQLite databases of version 3.0.0 or higher. The database schema is described in the software description.

MBTiles data cooperates with OpenStreetMap XML data and stores them in the relational database SQLite.

The specification is regulated under Mapbox In-cooperation, a private company dedicated to the map making in the public domain. 
\subsection{XML with DOM/JDOM and SAX}
Extensible Markup Language (XML) is one of the most common interfacing languages. It is a human-machine readable format, which makes it ideal to deliver graphical, structural information. The information carried in the XML file is contained in a tag and becomes an element in XML. The nested structure of tags makes the document organised from the root to its sub level, parents, and their child element\cite{St.Laurent2005}. 
\begin{figure}[H]
	\centering
	\includegraphics[scale=1]{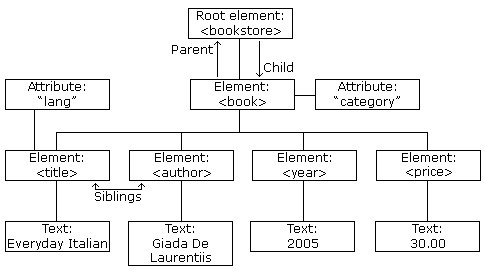}
	\caption{A DOM example of tree structure}
	\label{fig:dom}
\end{figure}
The Document Object Model (DOM) is a concept reflects the architecture of a XML file\cite{WHATWG2018}. It interprets the XML nested elements in a tree structure and read it from one node to another. It is a common technique to read through the XML file and locate the target of interest. An example showing a book store catalogue is shown in Figure \ref{fig:dom}

JDOM is the Java based DOM program, which helps the Java programmer to parse the XML file in the common known DOM method, hence saves time building a parser from the scratch.

\section{Map Renderer}
\begin{figure}[H]
	\centering
	\includegraphics[scale=0.71]{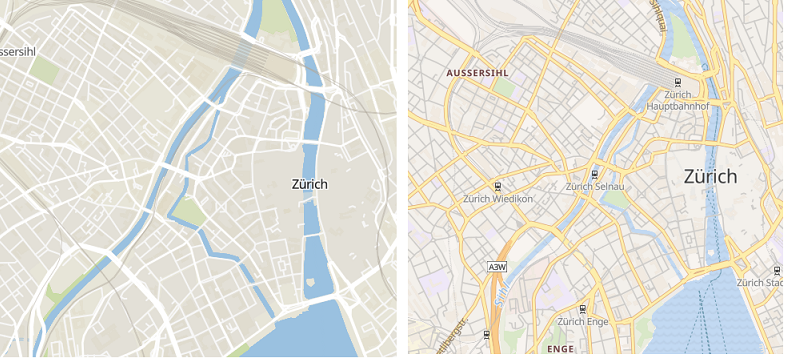}
	\caption{A same map rendered in two styles using stylesheets\cite{OpenMapTiles2018}}
	\label{fig:mapstyle}
\end{figure}
The map renderer reads the raw geospatial data from the database and then extract what it needs to produce a visual map, which contains blocks and colours on the screen. The map rendering requires the knowledge of computer graphics and geography. The data input is discrete information like polygon nodes and height. The render then outputs raster images or a set of raster tiles. It can also provide vector shapes that can be used through the map. The 3D object is generated based on the "level" or "height" information by default, and the style of rendering completely depends on render textiles library, though there might be some standards or guidelines.

The render helps the users to identify the object of interest in the most obvious way, visual identification. The interesting point can be exposed to the enormous amount of data and be used by developers. In Figure \ref{fig:mapstyle}, there are two different styles of the same map. These details and features are controllable in the style-sheets file. The JSON file acts as the agent to deliver the shape and colour information to the render.

The style of the map is defined in a separate file, which similar to the style-sheet found in the web pages

\section{Graphical User Interface for Geofencing software}
\subsection{Colour Scheme of the Geo-fence\label{sec:colour}}

\begin{figure}[H]
	\centering
	\includegraphics[scale=0.5]{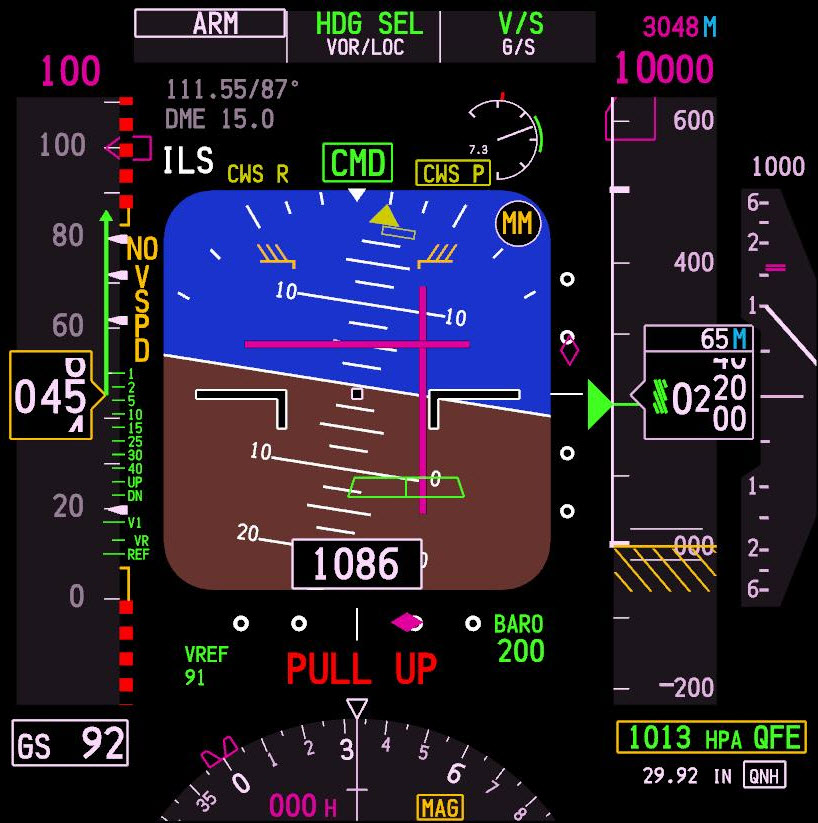}
	\caption{Primary Flight Display of Boeing 737 simulator\cite{PmFlightLimited}}
	\label{fig:pfd}
	
\end{figure}

The colour scheme is intended to indicate the rough height and the level of access for the UAV operator. With the multiple attempts of multiple layers of the map, the colour is displayed based on geofence decision. In order to have a functional colour display service, the geofence decision is used to put geofenced objects' raster image in different layers and in different colour. A typical usage of the coloured display is shown in Figure \ref{fig:pfd}

In Figure \ref{fig:pfd}, a typical aviation colour scheme is well demonstrated in it with following explanation. 

Magenta colour, in the middle and around the corners, shows a target for aircraft to reach, for instance, the purple box on the upper left sets a speed target on the speed-strip. 

Green colour, stands for a specific value setting, for example, the cursor on the right height slide and the BARO values surrounded by green boxes. 

White colour is used to show the information, from the speed to height, all numbers are marked in white colour and also applies to the artificial horizon in the middle panel. On Airbus platform, blue colour is also used for information display.

Red Colour is always for warning and dangerous. The upper side of the speed slide and "Pull Up" warning message in the middle are displayed as red. They either indicates the unsafe status or require immediate response is required.

Yellow or amber colour usually stands for minor failure or caution information for the flight crew, these alerts do not require instant response and should not have direct consequence to flight safety\cite{skybrary}.

The background of the display is set to be black, in order to minimise the effect of glared screen under direct sunshine\cite{Livada2012}.

\begin{figure}[H]
	\centering
	\includegraphics[scale=1]{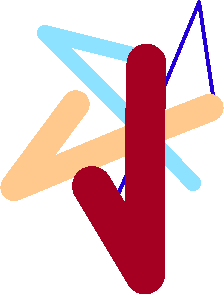}
	\caption{An example of multibands coloured raster}
	\label{fig:raster}
\end{figure}

The algorithm of putting colour on different raster layer is simple, shown in Figure \ref{fig:raster}, the different raster bands are overlapped with a colour scheme, which is defined in the colour map function of PostGIS. 

\subsection{XML/JSON as the graphical interface descriptive language}
\begin{figure}[H]
	\centering
	\includegraphics[scale=0.4]{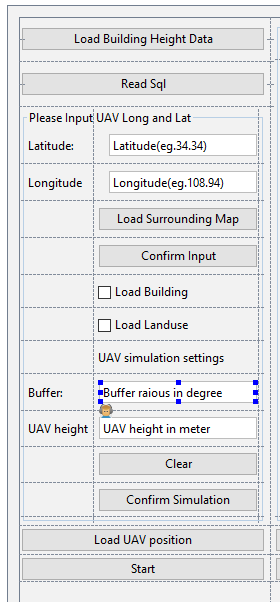}
	\caption{A text field in Java Swing GUI}
	\label{fig:guibuffer}
\end{figure}

Graphical User Interface(GUI) development is a tedious work without the visualisation development tools. XML is used as a descriptor for GUI frame. It enables the drag and drop tools to be used in the GUI design and retain the layout information in XML.
An example using XML to describe a Java GUI component is shown in Figure\ref{fig:guibuffer}, using the XML code in list \ref{code:guiexample}

\begin{lstlisting}[language=XML, caption=Java GUI Form textfield Example\label{code:guiexample}][H]
<component id="5e93e" class="javax.swing.JTextField" binding="bufferRaiousTextField">
<constraints>
<grid row="7" column="1" row-span="1" col-span="1" vsize-policy="0" hsize-policy="6" anchor="8" fill="1" indent="0" use-parent-layout="false">
<preferred-size width="150" height="-1"/>
</grid>
</constraints>
<properties>
<text value="Buffer raious in degree"/>
</properties>
</component>.
\end{lstlisting}

In the XML code in list \ref{code:guiexample}, a series of information, including row number in the frame, size and displayed text. From the XML code, a parer can convert it in to GUI whichever languages or platform the developer is using.
\chapter{Methodology\label{cp:3}}
As a software project, the development has to follow the pattern of software engineering. Not surprisingly, this project will not use Agile Development or any related concept, since the project is subject to be released in a short period of time with all functionalities. 
\section{Software Development Process}
 The model used is the V model, original from the Aerospace industry but in a simplified version. The V model is based on the waterfall model, which is the sequential development process of software engineering. The main sequence in the V model is the same as the waterfall model, but it emphasis the verification and validation throughout the development.
 \begin{figure}[H]
 	\centering
 	\includegraphics[scale=0.7]{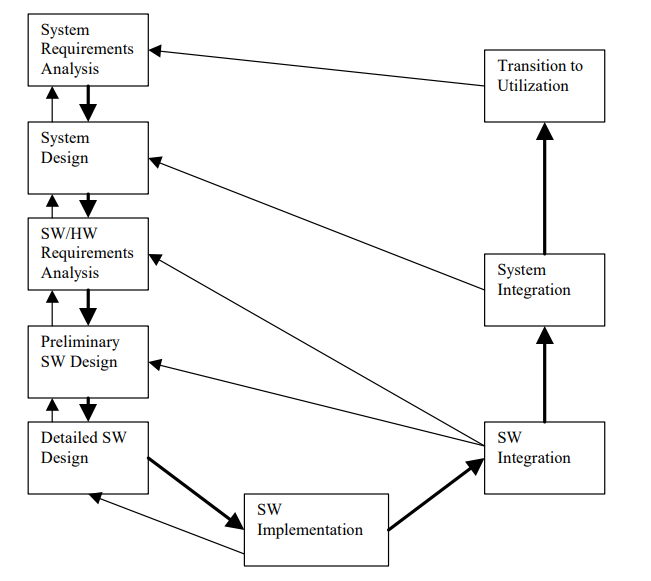}
 	\caption{The workflow of V-Model\cite{Johansson1999}}
 	\label{fig:vmodel}
 \end{figure}
In Figure \ref{fig:vmodel}, the workflow is closely connected with the integration process, and so is the protocol of Geofence Software development. 
\section{GAUSS project standards reference}
Meanwhile, due to the background of the Individual Research Project, the requirements are given and the safety assessment is removed from the model. 

This project does not conduct the safety assessment , since it severs as the system exploration rather than a complete life-cycle design. 

However, the safety regulations that adopted by this project are listed below:
\begin{itemize}
	\item CAP 393-Air Navigation Order 2016\cite{TheOfficeoftheGeneralCounsel2016}
	\subitem Article 241-endangering safety of any person or property
	\subitem Article 94A-small unmanned aircraft; height restrictions on flights
	\subitem Article 94B-small unmanned aircraft: restrictions on flights that are over or near aerodromes
	\item CAP 722-Unmanned Aircraft System Operations in UK Airspace-Guidance\cite{Group2001}
\end{itemize}
The project development surrounds the above listed standards and oblige to them.

\section{Software Deployment Consideration}
The Geofence Software, as an academic software prototype project, has no funding and resources for private content and data. Therefore, the use of open-sourced tools and public licensed tools is essential for the purpose of that Geofence software can be used under most non-commercial circumstances and not violate any copyrights and patents. Besides, the Geofence Software considers using well-established toolchain from its field, in this case, PostGIS and OpenStreetMap. The project shall be independent of any inclusive content or tools which cannot be obtained from the public internet access. The make of this software is subject to General Public License (GPL) licensing terms and may be distributed without the author's permission.

\subsection{The language of choice}
In this thesis, the major part of the program is written in JAVA with JDK 10. The choice of language is influenced by several reasons, but the efficiency comes first. Besides the Java language, several languages have come up to the choice but did not make their way to the project.

The Geofencing software requires a functioning Graphical User Interface to display the info of geofence. The needs of a Graphical User Interface can not be simply done in C/C++ language, where a comprehensive understanding of Computer graphics is required. In fact, building a GUI in C/C++ is not only troubling the development but also the deployment of the program, since the compatibility of C/C++ GUI is pretty Operating System specific. The cross-platform GUI requires knowledge of external, 3rd party software, such as Qt, to be involved. In this case, the plan to design and develop using C/C++ is ditched. 

Python is among one of the most popular languages these days, not only because of user-friendly philosophy but of its wide usage in machine learning and data analysis. Python has a great advantage to be chosen as the project main language, however, the performance has undermined its application on the drones, especially, in the real-time environment.
The way python works are based on the interpreter, and it causes significant performance drop, not to mention it reluctantly persists concurrent single-threaded execution.

Java\texttrademark is the final choice of the language due to its perfect balance of performance and developing efficiency. The GUI is constructed under JAVA Swing API, a comprehensive set of graphics libraries that enables all kinds of interaction required by the geofence project. 

The object-oriented design facilitates the organisation of the program and simplifies the interface internally. The potable compilation can be used in any operating system that has Java Virtual Machine runtime, therefore provides great cross-platform dependability. These are the main reasons for the choice of JAVA.

\subsection{Database of choice}
Database Management System (DBMS) is the essential part of the Geofence software, it plays the role of manipulating and storing the spatial data. This project uses PostgreSQL as the main database, as the PostGIS extension is the most suitable tool for Geofence. In addition to this, the ethos of open source is also a driving factor, let along the high-performance engine comparing with other comparative DBMS. It also has a very good compatibility with JAVA using JDBC library and works on most operating systems including Windows and Linux.

\begin{figure}
	\includegraphics[scale=0.25]{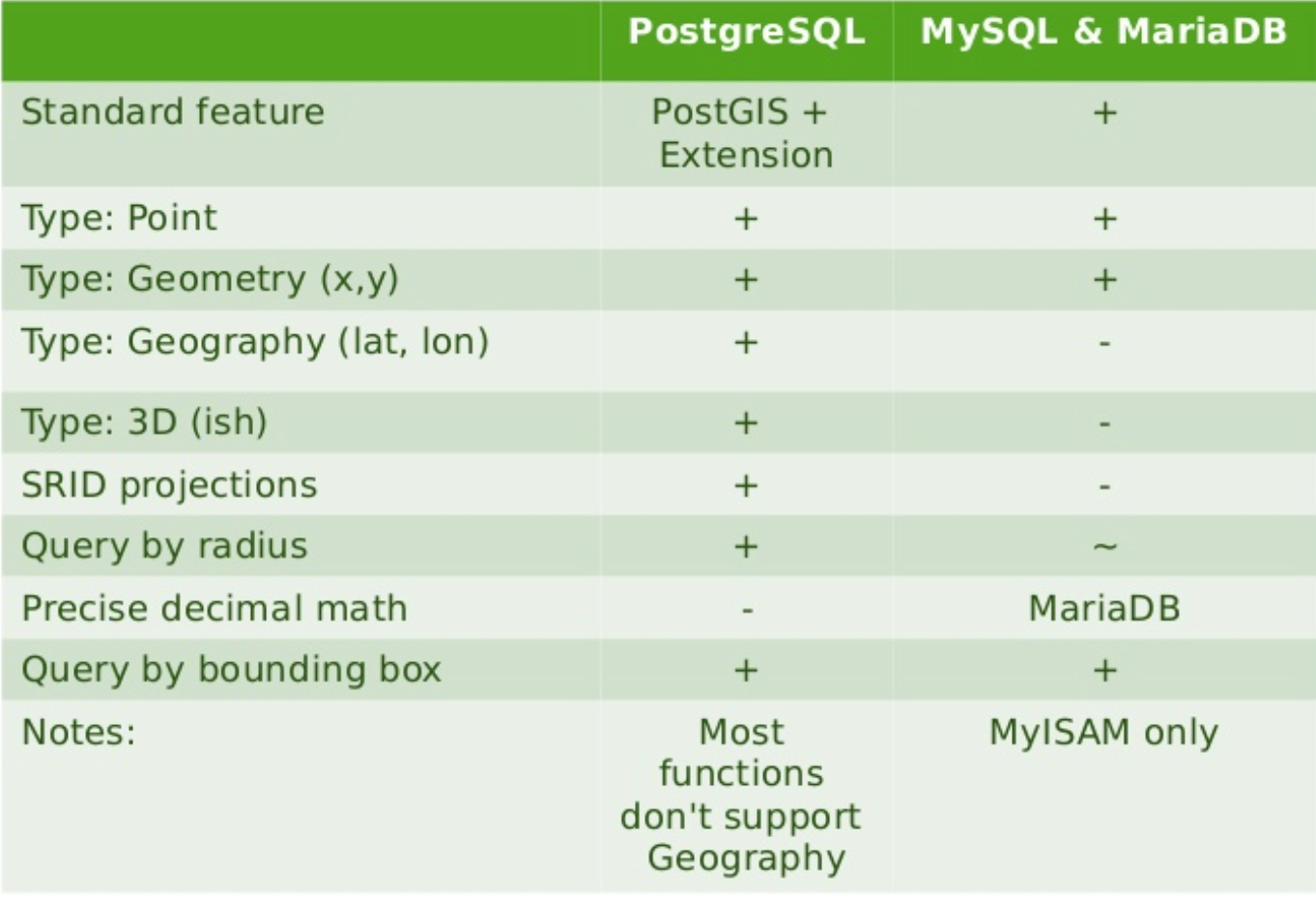}
	\caption{Comparison of PostGIS and MySQL database system\cite{Ingo2013}}
	\label{fig:postmysql}
\end{figure}

\begin{figure}
	\includegraphics[scale=0.7]{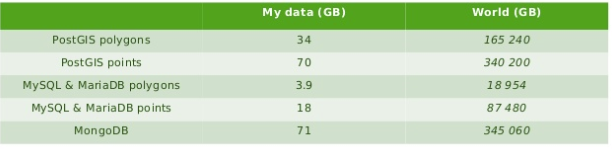}
	\caption{Data size comparison among databases.\cite{Ingo2013}}
	\label{fig:size}
\end{figure}

The main competitor is MySQL DBMS, which is also open-sourced and well-established. In Figure \ref{fig:postmysql}, numbers of databases are compared, and shows that MySQL lacks some functions. In Figure \ref{fig:size}, a comparison of data size of polygons and points data shows that MySQL has the best storage efficiency.  However, lack of spatial data tools and spatial function extension has limited its usage in this project.
Considering all the conditions, the PostgreSQL is the only choice for this project and will be used as the backbone from beginning to end.

\chapter{Geo-fence Software Architecture and Function Design \label{cp:arch}}
The Geofence software consists of a map loader, geofence generator, geofence advisory program and geofence display render.

\section{Architecture Overview}
The architecture of Geofence software is designed based on all the needs and hardware platform availability. The hardware is limited to the general x86 machine, where the performance should not be a concern. The machine is capable of running Java Virtual Machine and PostgreSQL database.  
Generally speaking, the architecture should not only support all the functionalities of the software but also pay attention to performance, availability, scalability, expandability and security. A good architecture should balance the factors all the directions and focus on the main target, functionality. It is not a good architecture if the intended function is not functioning.  Clearly, the scope of Geofence software is not covering deep about software development, neither focusing on software architecture design orientation, still, it is something important to the development. 

There are two databases in the main architecture for different purposes. The PostgreSQL database with PostGIS extension offers the geofence data processing and advising functions to the main program and the render uses SQLite database to process rendered tiles in order to provide rendered images of the geofence.

\begin{figure}[H]
	\includegraphics[scale=0.85,angle=90]{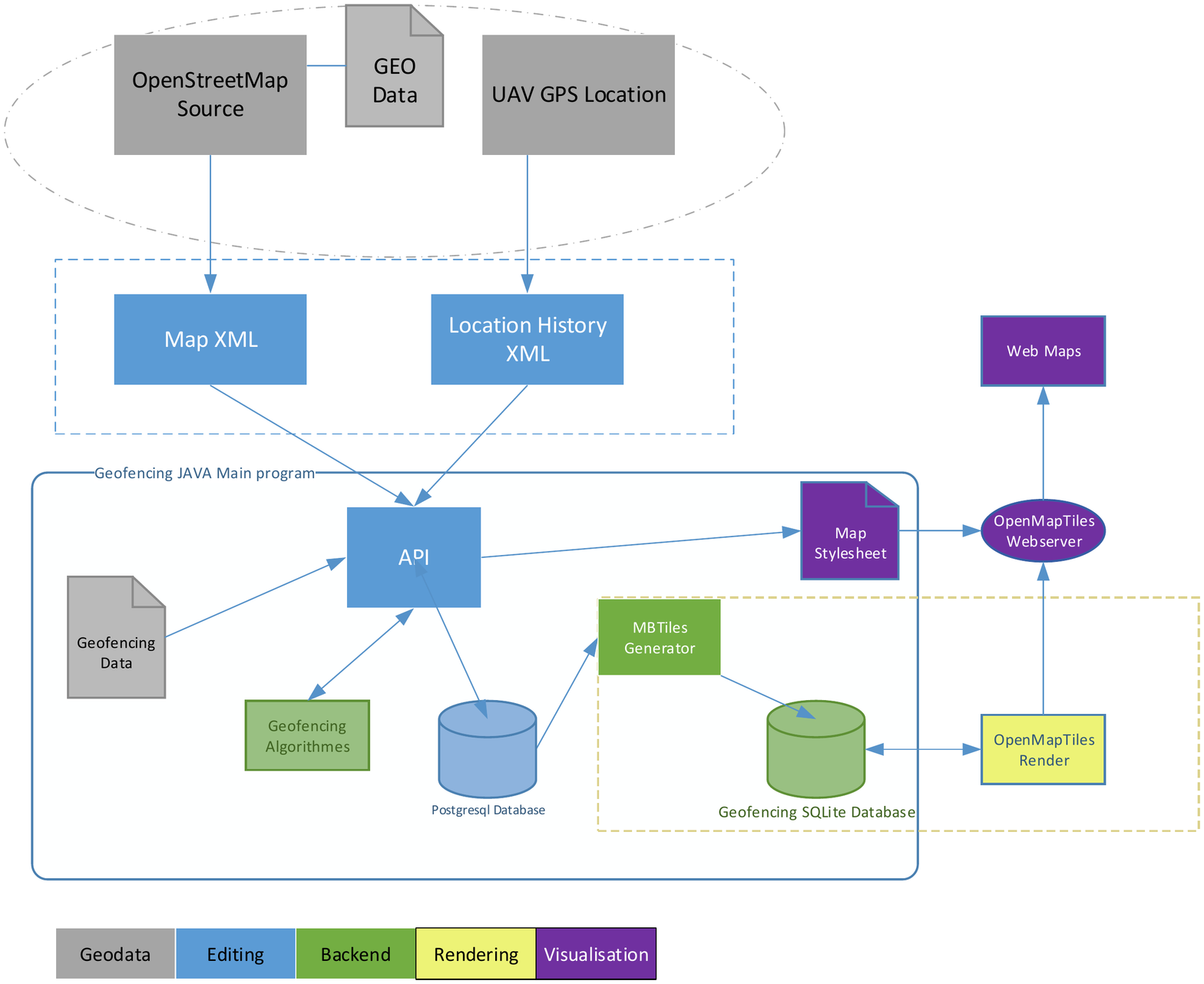}
	\caption{The purposed architecture of the Geofence Software}
	\label{fig:arch1}
\end{figure}

In this project, the OSM website architecture is the template for creating Geofence software, as can be expected, with comprehensive modification and adaptation of new purposes. Here, in Figure \ref{fig:arch1}, the mainframe is designed with mostly similar components as the OpenStreetMap architecture, which is shown in Figure \ref{fig:osm}, the similarity between these two architectures are caused by website oriented design. 

It uses similar function blocks and tools to achieve rendering and editing. This purposed architecture has simplification on the data source and map render, where the external render is introduced. It also removes the external APIs, for example, the OVERPASS API and reduces the layers of the map visualisation. In brief, a down-sized OSM architecture with only the essential components. It uses OpenMapTiles server solution to replace the complicated tile rendering functions and offers great visualisation in a Web-based environment. It is indeed a good architecture based on the assessment of functions and other factors. Nevertheless, the web visualisation is an extra part of the Software and requires more computing resources to handle, correspondingly, the operator needs to pay attention to Map visuals and UAV status. The already made OpenMapTile server is hard to modify as it is packed in a docker container and requires the docker to deploy. In conclusion, it is a well-balanced architecture but lacks flexibility.
\section{Architecture for Geofence Software}
\begin{figure}[H]
	\includegraphics[scale=0.85,angle=90]{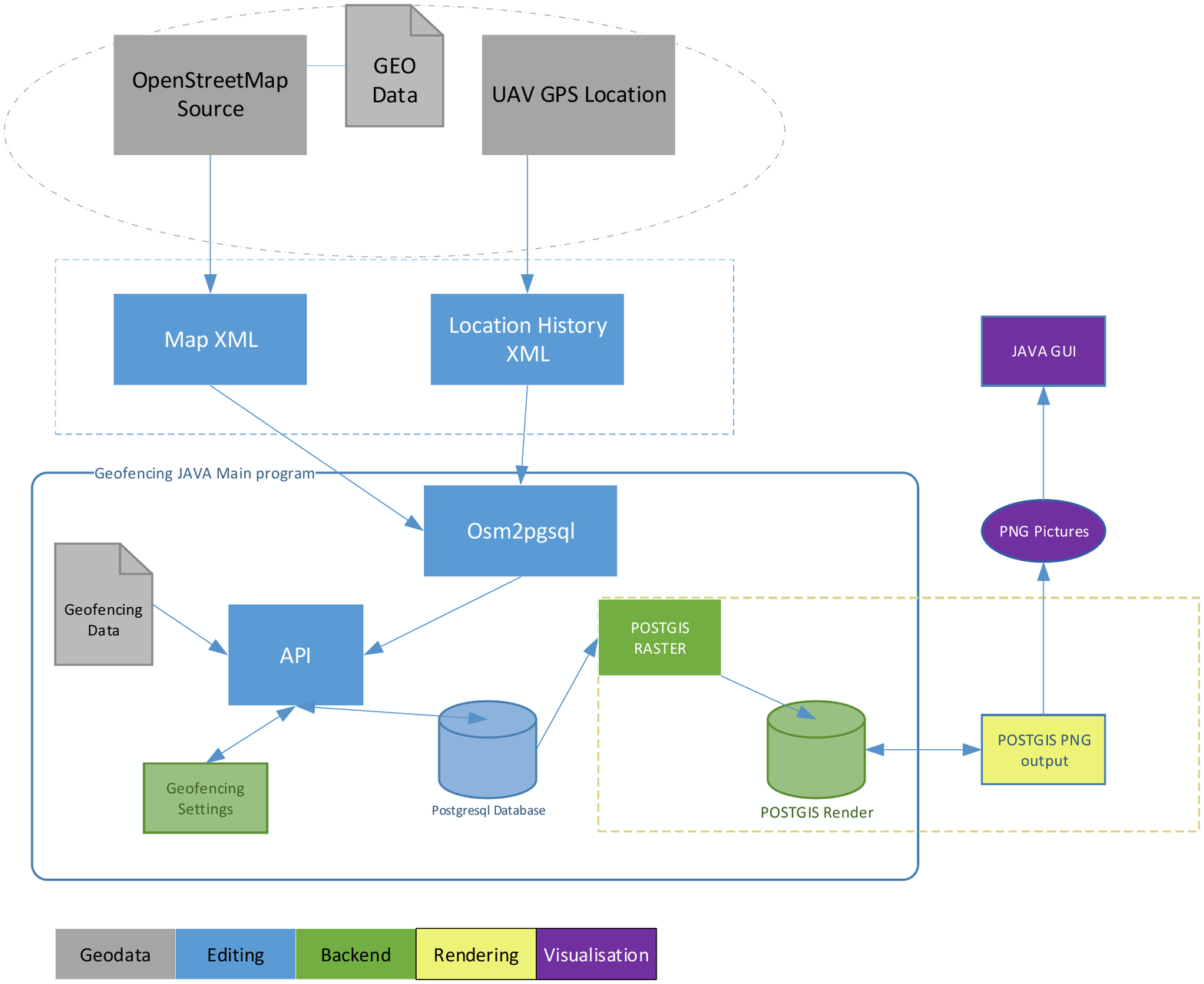}
	\caption{The JAVA GUI led Architecture}
	\label{fig:arch2}
\end{figure}

To counter the drawbacks of the purposed architecture, the architecture is shown in Figure \ref{fig:arch2} compensates the flexibility and integrity of the software, using all built-in tools and renderer inside the PostGIS. Of course, it sacrifices the beautiful GUI of map visualisation and the docker container feature. In contrast to the Web-based map, the new architecture uses local JAVA GUI to display images rendered by the PostGIS with PostgreSQL database.

Benefit from the optimisation, the database is then unified to PostgreSQL only and removes all the external dependency. It is now self-contained software with very high-level integration.

\section{Geofence software auxiliary functions design}

\subsection{Map Loader}
This function provides the interface to interact with Map Sources, for example, OSM and OS Map. These map providers have their customised map form and different reference model, so the unification of the maps is important to the geofence. It is part of PostgreSQL stack rather than the built-in function of the Geofence Software.

During the conversion of all sorts of map data into the database, the SRID of the map has to agree. Different SRID may result incompatible data and cause errors. The default SRID in this project is 4326, which stands for WGS84.

The map file should be named as "map.xml" for the Geofence to import. The purpose of having the redundant XML parser is for externally map source. It supports all OSM compatible XML data. 

\subsection{Height Importer}
This module is designed to import object height information from the reliable source. However, the source of height information currently has not met the standard for the geofence. Hence, the reserved function leaves the import port once the source has reached enough precision and reliability.

This function allows import of height from XML documents and offers an easy to use interface to the operator.

The XML format Digital Terrain Model is available from OpenStreetMap's partner Hochschule für Technik Stuttgart. They have converted the SRTM mission data into OSM XML format and made it available to the public for free. It can be easily imported by the Height Importer.

On the other hand, the Building height data is highly independent of public resources, as far as it is concerned, the public survey data does not include any building height information but a private survey. It is a bit tricky to obtain them under this project premises, however, the data is possible to be used without any extra effort. 

\subsection{Geofence Construction File}
The Geofence is defined by parameters of the objects in the map. However, the cached objects are the bearer of the geofence. Therefore, the size of the cached area shall be selected by the user, depending on the UAV flying velocity. Beside the buffer size, the Geofence parameter shall also include obstacles from any known objects type. 

These details are the construction file, which should be defined by the user. The Geofence Software will read the configuration from the construction file and performing tasks accordingly. 

\begin{figure}[H]
	\centering
	\includegraphics[scale=0.7]{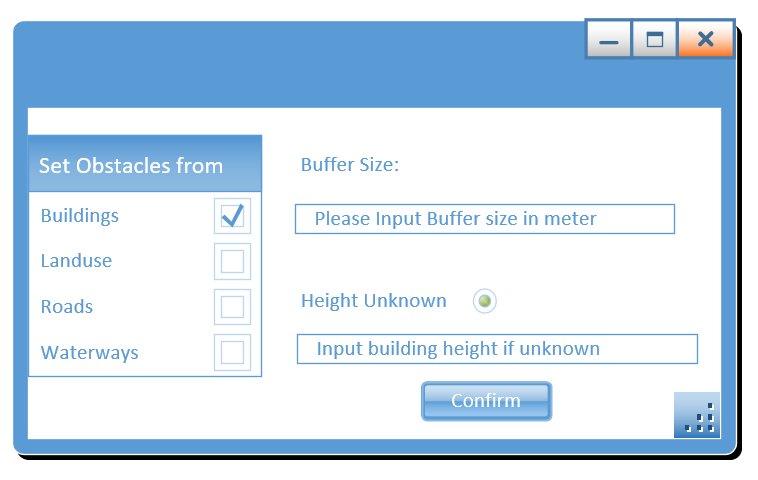}
	\caption{Construction File Helper Design}
	\label{fig:const}
\end{figure}

The above parameters are editable via the simulator interface, and a dedicated graphical interface is planned for future, as shown in Figure \ref{fig:const}. The helper program will generate the file for users for the use in Geofence Software.

\section{UAV input and output design}
The UAV input is not specific to any type of methods, which means it is possible for all kinds of possible input. The intended solution is to convert them to string or plain text in the machines file system. Hence the file system and the Geofence has an interface, which is unified no matter what input it is. The input and output will be handled in the same fashion and the further work of using the plain text will be up to the UAV controller designers' hand.

It is recommended to use JSON or XML as the interface file and should be included in the future works.
\section{Geofence Display design}

The Geofence Display comes with two versions, a simulation program and a UAV hand-held version. The simulation program follows general guideline of a desktop software and the UAV hand-held version uses completely different set of rules.

The colour scheme used in the UAV hand-held terminal, follows general aviation colour convention mentioned in section \ref{sec:colour}, Chapter \ref{cp:2}.

Not only the obstacles image, the reference map shall also be included for pilots reference. The reference geometries shall have different colour, in order to be identified from the display.

\subsection{Geofence display sizing and control design}
The GAUSS project plans to deploy the Geofence software on hand-held device, with a palm size screen. Meanwhile, the hand-held device is attached to a mechanical control broad, which has controllers and switches on it. The display size of a typical hand-held UAV terminal is around 7 inch with 800 x 480 pixels. The tough screen option is good for Geofence Software but not compulsory.

\section{Geofencing Consideration}
Geofencing algorithm is the core of the Geofence Software. In the development, there are a few considerations.
\subsection{Geofence Rules making}
The Geofence obstacles are generated based on geofencing rules. The parameters can be used to classify objects into obstacles vary from type to type. In the basic map supply, there are five major categories : building, natural, landuse, roads and waterways. The algorithm can go deep into each of them and read the geometry and other attributes, such as, name, type, height and the osm\underline{  }id, if OSM data is used.

The Geofence software shall allow users to use any of the parameters as a filter to create obstacles lists. The obstacle will have restriction on entry. Meanwhile, users can decide which area has special permission of access, and a different list can be set up to contains these special objects to allow Geofence Software render them on display.

Building height information is supported by the Geofence software, however, the rule of making geofence using height is not clear at this stage. The regulator may want to limits the UAV flying over buildings with certain height. This is an open area of rule making.

In this practice of Geofencing, all buildings are assumed at 30 meter high and non-flyable for the UAV.
\subsection{Buffer size decision}
From the UAV classification, only group 1 and group 2 are considered as the Geofence Application Users, therefore the buffer size is decided according to the maximum speed of the UAV in general, where the special buffer can be defined individually.

\begin{figure}
	\centering
	\includegraphics[scale=0.4]{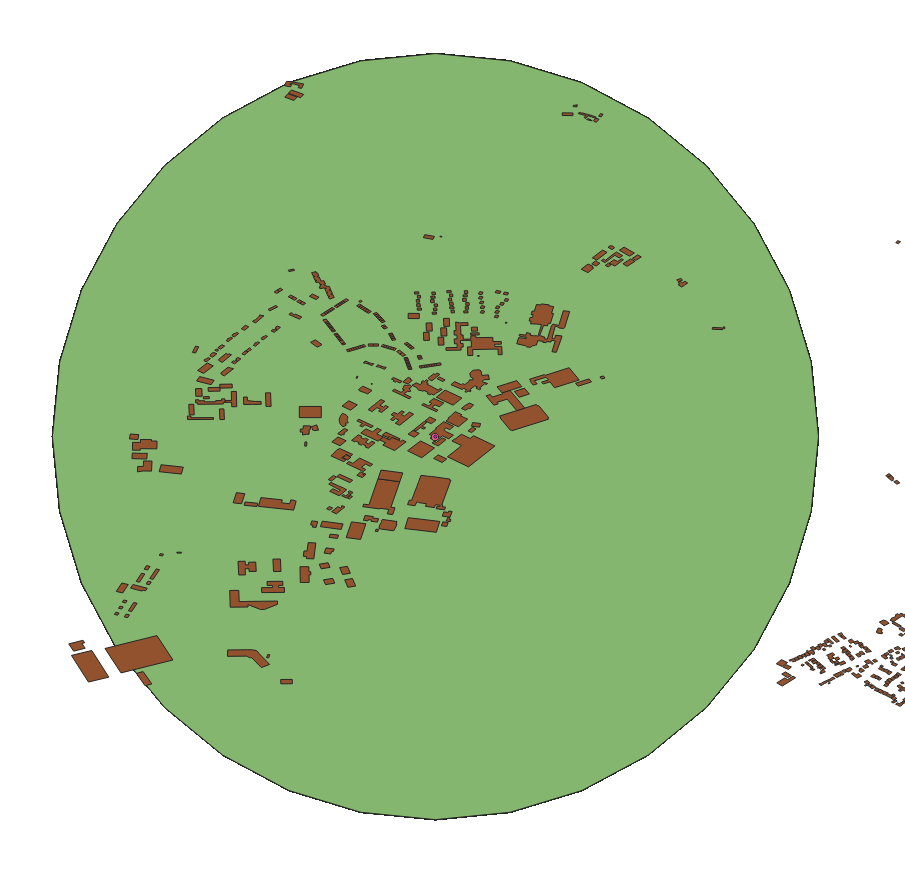}
	\caption{A 0.012 degree circle over Cranfield University}
	\label{fig:buffer012}
\end{figure}

As the maximum speed in above classes is 250 knot, which is 128.6 meter per second. To leave enough time for decision making, 20 seconds results in a buffer zone raids of 2572 m and 15 seconds yields 1950m. The size of the rendered map is essential for the operator, as the size of the building varies according to the buffered zone. An example showing a buffered circle of 0.012 degree hovering on the Cranfield University, is illustrated in Figure \ref{fig:buffer012}

Most commercial UAVs, flew by unlicensed operators, are low-speed vehicles comparing with large-scale fixed wings UAVs, are quadcopters. The typical speed of them is around 60m/s, therefore, the chosen buffer size is around 1000 meter, which will be defined by a degree in the WGS84 system. 

Using the ST\underline{  }transform function provided by PostGIS, a 1044.5 meter radius circle is projected as 0.012 degree radius in WGS84, and the resulting area of the circle is 3427475 meter square. This circle size will affect the performance of the Geofence. 

The dynamic range of buffer size is under demand and may come in further updates.
\begin{figure}[H]
	\includegraphics[scale=1]{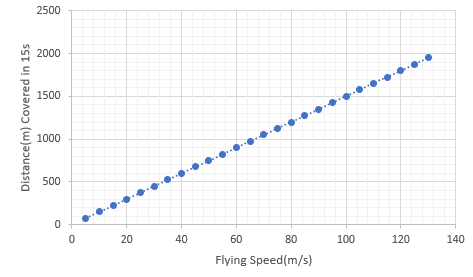}
	\caption{Buffer size against flying speed}
	\label{fig:buffer}
\end{figure}

\begin{table}[]
	\caption{Distance per degree of longitude }
	\begin{tabular}{|c|c|c|c|c|}
		\hline
		\multirow{2}{*}{Difference of longitudes} & \multicolumn{4}{c|}{Actual distances}                      \\ \cline{2-5} 
		& at 0$\deg$ lat.    & at 30$\deg$ lat.  & at 60$\deg$ lat. & at 87.5$\deg$ lat. \\ \hline
		0.01 $\deg$                                    & $\sim$1 000 m & $\sim$870 m  & $\sim$500 m & $\sim$43.62 m \\ \hline
		0.001 $\deg$                                   & $\sim$100 m   & $\sim$87 m   & $\sim$50 m  & $\sim$4.36 m  \\ \hline
		0.000 1 $\deg$                                 & $\sim$10 m    & $\sim$8.7 m  & $\sim$5 m   & $\sim$0.44 m  \\ \hline
		0.000 01 $\deg$                                & $\sim$1 m     & $\sim$0.87 m & $\sim$0.5 m & $\sim$0.04 m  \\ \hline
	\end{tabular} 

\end{table}

\subsection{Geofence Advisory}
Advisory works to provide alternative route for intended flying course. The way to Figure out the intended course is based on the velocity and heading model. With no prior knowledge to flying plan or navigation information, the advisory function shall provide diverting route via the text panel in the GUI or in command line, ideally, the guidance message, which is generated by the Geofence Software, shall be fed into the controller, and divert the aircraft automatically. 

Due to the low integration level of this project and ongoing flying control system project, the direct fed is not included in the scope. However, when needed, the message can be called or accessed from output stream of the Geofence Software Output.

The diversion shall be considered together with way-point system, which may be used by the operator to plan the route. The Geofence Software will read the next way-point, and virtually construct a line between the UAV and the next way-point, then judge if the line intersects with the genfence in the database. The result from Geofence will indicate the operator the diversion is needed.

In addition to the text advisory system, an audio warning message will be played in the event of approaching genfenced object. The recorded voice will tell user to hold the aircraft to prevent fence violation. 

The audio alert is not presentable in the thesis, but included in the code submission.

\begin{figure}[H]
	\centering
	\includegraphics[scale=0.5]{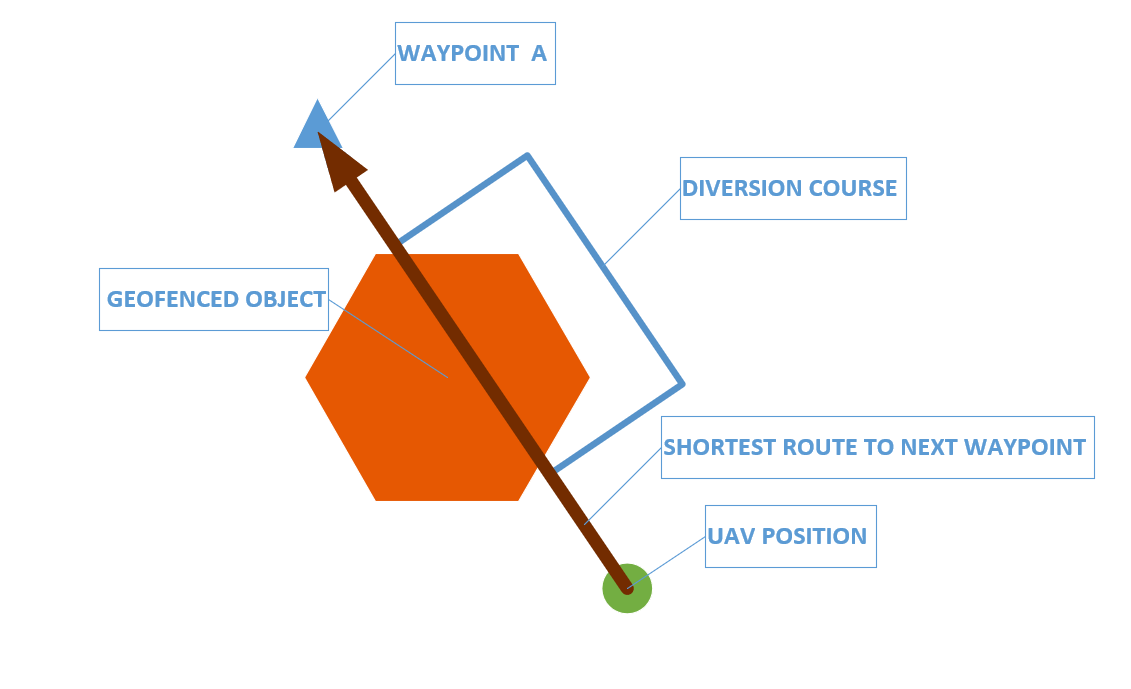}
	\caption{Divert course by the Geofence Advisory}
	\label{fig:ad2}
\end{figure}

In Figure \ref{fig:ad2}, the next way-point is present, and the geofence object lays in the middle of the UAV and the way-point A. The Geofence Software will advise the operator to take divert course. In this release of the Geofence, the divert course calculation is not included as it requires more UAV operating experience and knowledge, which is out of the scope.

\begin{figure}
	\centering
	\includegraphics[scale=0.6]{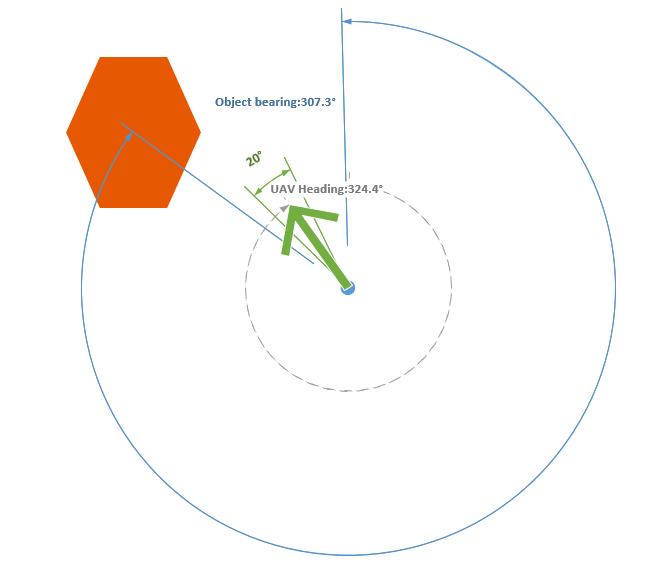}
	\caption{Example of decision making}
	\label{fig:ad3}
\end{figure}

To determine whether an advise is needed, the Geofence Software will compare the UAV heading and the object's bearing from the UAV. If the object's bearing is within 10 degree range from the heading, a message from the Geofence Advisory will be delivered to the operator, an example in Figure \ref{fig:ad3}, shows a UAV is heading to 324.4 degree and the object's centre is at 307.3 degree from the UAV, the difference is 17 degree, so that Geofence will not produce warning message hence the divert is not required.

The degree decision, however, is not perfect, in terms of the geometry, because the rest of the object may exceed the range of 10 degree, and may still stay on the course. But the truth is, if the object is that big, the operator shall be able to tell by him or herself. It is now an intermediate solution, where the improvement can be made.

Either way, the decision-making requires human input. The diversion route is never easy to make, and the safety is the prior concern. In order to achieve ultimate goal of geofence, by keeping the human in the loop and advisory only acts as advice, the advisory will improve the geofence awareness and help operators fly with UAVs.

\subsection{Geofence Situation}
\begin{figure}[H]
	\centering
	\includegraphics[scale=1]{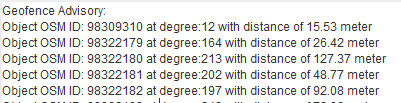}
	\caption{The advisory information example}
	\label{fig:ad}
\end{figure}

Geofence Situation indicates the surrounding geofence to the operator, for navigation and guidance purpose. The information from the software includes the surrounding object's osm\underline{  }id(if there is one), degree of angle from the UAV and the distance to the geofenced object. A sample from the running Geofence software simulation is shown in Figure \ref{fig:ad}.

In this example, the osm\underline{  }id is displaced instead of the name. The reason is simple, the object is not always named, for instance, buildings in Cranfield University. The user can supply a more comprehensive database of objects, then it can be updated to display the names.

The text advisory will not be displayed to the graphical user interface but stored in the log file for further procession.

\begin{figure}[H]
	\centering
	\includegraphics[scale=1]{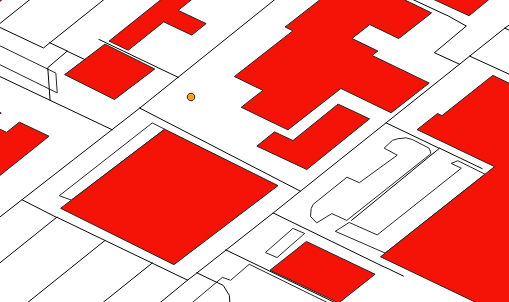}
	\caption{Example of a reference map interface with orange dot representing UAV location}
	\label{fig:situation}
\end{figure}

The graphical user interface will display the fence objected in the red colour and the rest of the map will be rendered in white, as shown in Figure \ref{fig:situation}

\subsection{2D, 2.5D and 3D Geofence}
As initially planned, a geofence should work within the real world, where the dimension is 3. However, regarding the most geofence requirement and regulations, the 3D geofence is more likely a 2.5D object due to the reason that the object is not 3D modelled but a 2D polygon shape with a height information.

A 2D Geofence is a polygon or multi-polygon geometry that restricts the entry or exit of the vehicle only by its planar coordinates, for instance, latitude and longitude. It is not enough for a flying object, such as UAVs. The Geofence should be capable of dealing with altitude change during the motion and prevent entry or exit above or below a certain height. Hence a 2.5D geofence is constructed using a 2D polygon geometry and a height. The height importer will link two tables together with unique id, in this case, the osm\underline{   }id, as the reference. It will construct a relationship but not a 3D object, which also reduces the size of the data. The performance is improved by introducing this feature.

\begin{figure}[H]
	\includegraphics[scale=0.6]{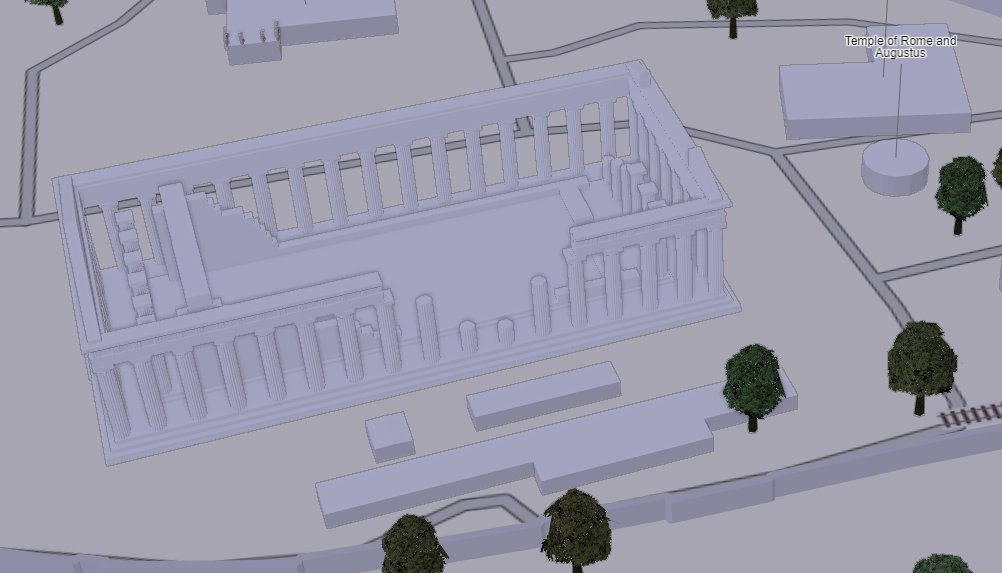}
	\caption{A example of mixed models \cite{OpenStreetMapContributors2017}}
	\label{fig:model}
\end{figure}

The 3D geofence is particularly useful if the high precision geofencing in presence. The 3D modelled object can have great detailed features such as triangle roofs, sloped walls and erected poles. These features will help UAVs to fly safely. However, in most modelled maps, the models of buildings are combinations of simple 2.5D made cubes and the well-made 3D models. An example of two types of buildings models are shown in Figure \ref{fig:model}

In this example, the Parthenon's model is a 3D model based on its real-world appearance, however, the camp building in front of it is merely a rectangular shaped box. The effort of making good models in the level of the Parthenon is way harder than the camp building.


\section{Geofence Performance}
Geofence Performance measures the time consumption of making the fence and producing advisory information.
The Performance test is consist of two parts, the Java application performance and the database response time.

The first part will examine the performance of built Java Jar artefact on the target machine. It uses Java Virtual Machine Profiler to analyse the performance and resource usages, such as, CPU utilisation, thread status and time consumption.

The second part of the test is conducted inside the database, by setting different buffer size, the response time of raster output and fence generation can be obtained. It is a quantitative result set of comparing different parameter in the configuration.

\section{Geofence Work Flow design}
\begin{figure}[H]
		\centering
	\includegraphics[scale=1]{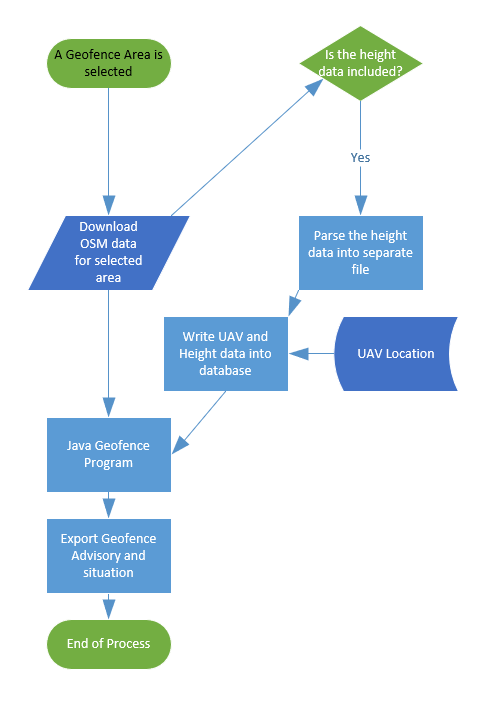}
	\caption{Work Flow of the Geofence }
	\label{fig:workflow}
\end{figure}

The Geofence software is design to work as a part of the work flow, the whole process is shown in Figure \ref{fig:workflow} and the software does not work alone if the required data is not prepared.

The major work of the Geofence Software includes Data importing, Geofence Generation, Geofence Situation Display, Geofence Advisory, Geofence information exporting. However, the core part of the Geofence project is all about Geofence Generation, Display and Export. Interfaces of import and export are described in section \ref{sec:interface}, where the detailed flow charts can be found.

The detailed program flow charts are dedicated in the Chapter \ref{cp:5dev} Section \ref{sec:class}

\chapter{Geofence Software Development\label{cp:5dev}}
\section{Interface Development}
Software interface has multiple layers content, for instance, the database interface, process interface and most importantly, Application Programming Interface(API).

The Application Programming Interface is a collection of methods to communicate and interacts with various parts of the software internally and externally. API is often designed to be used as building blocks, which can be easily called and assembled by fellow programmers. It is an extremely convenient tool to access object-oriented software. The notable example of API is the Portable Operating System Interface (POSIX), in terms of the operating system, and XML, as an interfacing language.
\begin{figure}[H]
	\centering
	\includegraphics[scale=0.6]{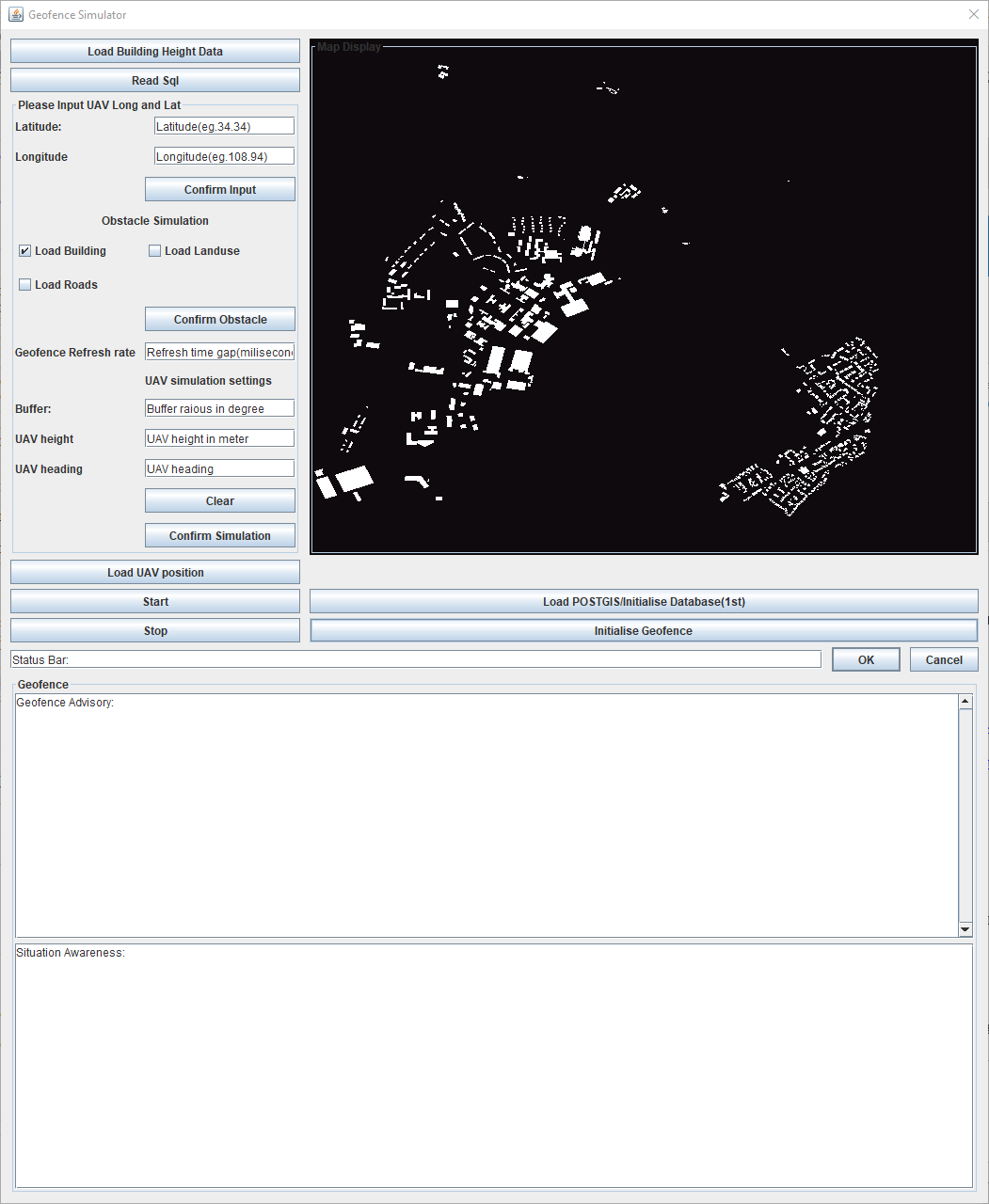}
	\caption{The User Interface Design Of the Geofence Simulator}
	\label{fig:ui}
\end{figure}
\subsection{Graphical User Interface}

The Geofence can operate without any human intervention by putting the pre-configured program in the target device, works as the background service to facilitate other services, for example, feed guidance message to flight control system. Meanwhile, Graphical User Interfaces are offered too.
The Graphical User Interface is designed for desktop and the hand-held devices with input and display hardware. The desktop version is the Geofence Simulator.
A design drawing of the simulator GUI can be found in Figure \ref{fig:ui} with a large buffered map display.

The Simulator UI design reflects the functions and interfaces of the simulation program, however, in the hand-held device, the parameters including UAV location, velocity and heading, will not be supplied by the operator but the UAV. 

The mobile version is a working Geofence software. The default resolution for the hand-held version is 800 x 480 pixels. The map size is 600 x 480 pixels and the rest of area belongs to the Geofence Advisory and UAV status.

The Geofence UI is the outcome of the JAVA GUI components and the runtime on the machine. It is designed with touchscreen philosophy, but also accepts mouse and keyboard inputs. The UI for UAV display matches the Aviation colour code.
\begin{figure}[H]
	\centering
	\includegraphics[scale=0.6]{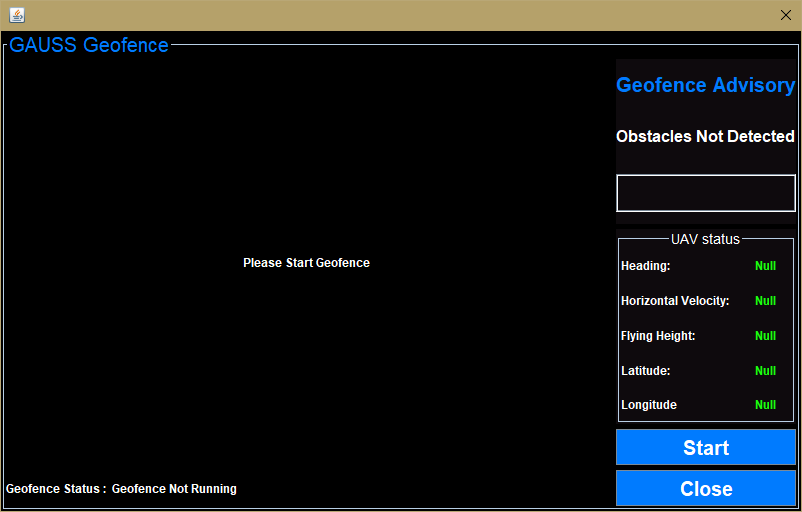}
	\caption{GUI design for UAV Hand-held Device}
	\label{fig:gui}
\end{figure}

Figure \ref{fig:gui} shows a design plot of the UAV hand-held device display. This design features the Geofence Display and also the Geofence Advisory functions. The left side is dominant by the Geofence Map and situation display, and the right panel is divided into three sections, Advisory, UAV status and a concise setting menu. 

The colour and font are designed for outdoor uses, with high contrast display and avionics general colour scheme. The pilot should be altered if the geofenced object is in appearance and the font will turn red.

Start button is for UAV operator to decide when to activate the Geofence and the Close button does it says, stop the program.

\subsection{Software Interface\label{sec:interface}}
\begin{figure}[H]
		\centering
	\includegraphics[scale=0.7]{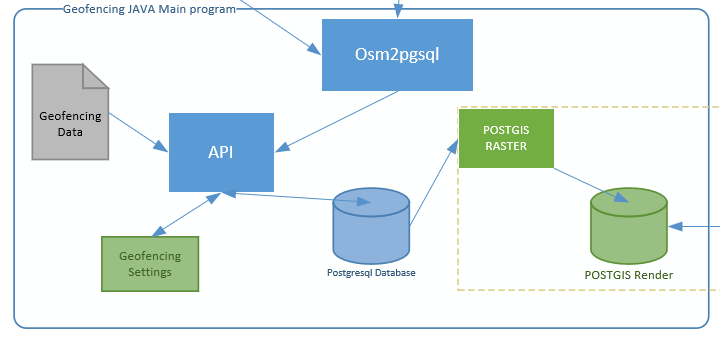}
	\caption{Major interfaces of Geofence Software}
	\label{fig:interface}
\end{figure}

The geofence needs to interact with users, maps, geofence advisory and UAV location data. They require a set of the exposed interface to do so.  In this project, the main interfacing is about the main Java program with rest of the parts. As shown in Figure \ref{fig:interface},

\subsubsection{Map to Geo-Database Interface}
\begin{figure}[H]
	\includegraphics[scale=1]{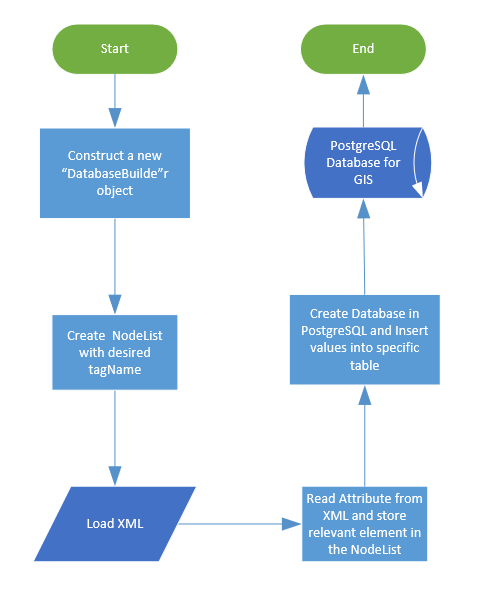}
	\caption{Flow Chart: Load XML to database}
	\label{fig:loadsql}
\end{figure}

\begin{figure}[H]
	\centering
	\includegraphics[scale=0.7]{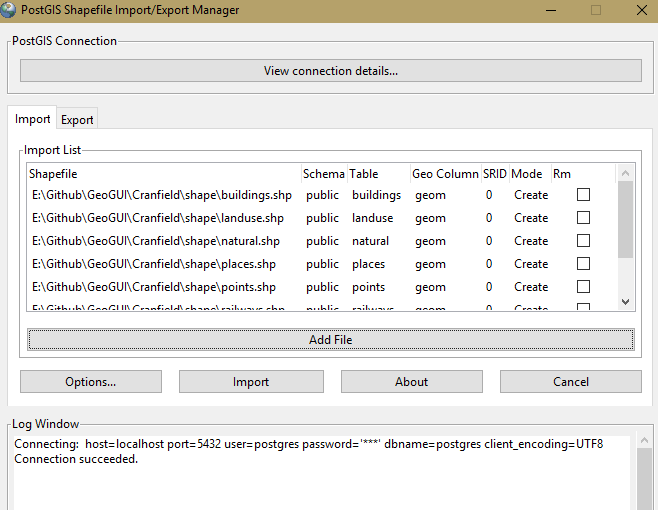}
	\caption{Loading shapefiles into PostgreSQL}
	\label{fig:loader}
\end{figure}

The data comes in XML and database, requires special API to deal with. The XML is parsed by the Java Class "DatabaseBuilder", with method"Load SQL". Detail work flow is shown in Figure \ref{fig:loadsql}. In this method, the XML file is parsed into the database using tag names. The detailed classification of the data is not discussed due to content scope issue.
The shapefile is welcomed in this Project, because the PostGIS current build contains a handful tool "PostGIS 2.0 shapefile and DBF Loader Exporter". It provides a GUI for user to select map they want, and import them directly. Figure \ref{fig:loader} shows the example of importing cranfield area map into PostgreSQL.

Besides, the map may come in as tiles, in this case, the map needs to de decomposed into polygons using PostGIS functions. This is not covered in this project, users who hold tiles data shall look for other open sourced tools for help.
\subsubsection{Geo-Database to Map Render Interface}
All the GIS information is handled by the Database, to use the information, the entry shall be exported or ready to be collected by the external render or user. The interface between each other is the Java Database Connectivity (JDBC) API. 

To make the raster, following flowchart shows an example to convert the geometry into raster.

\begin{figure}[H]
	\centering
	\includegraphics[scale=1]{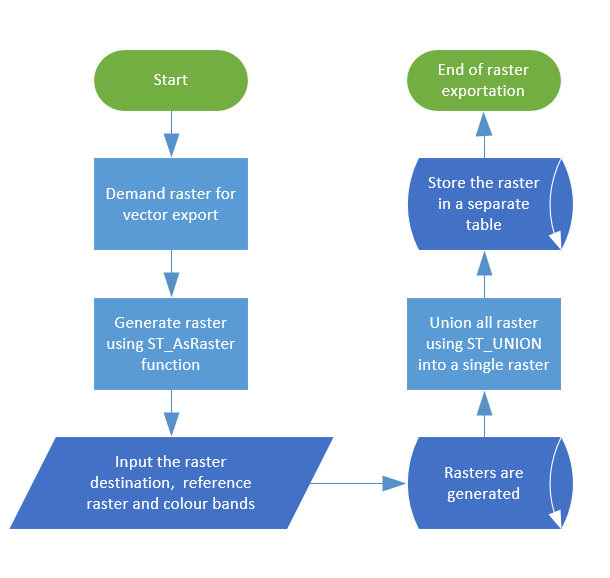}
	\caption{Flowchart of making raster from geometry}
\end{figure}

Meanwhile, if other programs are used, the Open Database Connectivity API shall allow users access record from any framework, any platforms. 

Interfacing file between the database is the raster data of the area. It can be exported as any image format or be further processed. To convert a geometry to a raster, additional information like colours and size is required, and for the sake of consistency, a reference raster is recommended, for example, the original map raster or a empty raster with correct position. If the reference raster is not referred, the misalignment may occur while overlaying multiple rasters. 

\subsubsection{Render to Visualisation Interface}
\begin{figure}[H]
	\centering
	\includegraphics[scale=0.7]{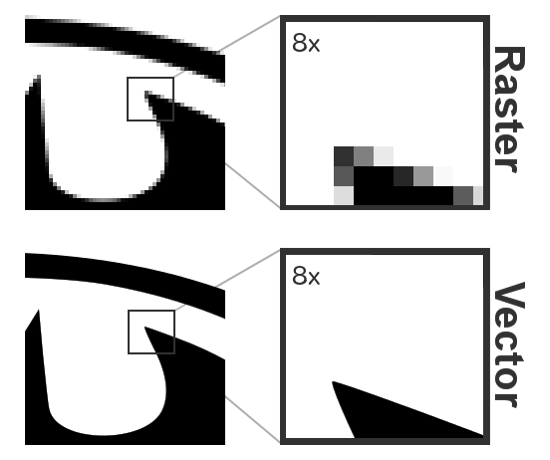}
	\caption{Compare Raster and Vector\cite{ArcGIS2018}}
	\label{fig:raster2vector}
\end{figure}

The Modern GIS software handles multiple format of spatial data and exports them to every-day software, as pictures or map tiles packed in a compressed file. The effort extracting images from the Geofence Software is by converting raster data into vectors and then vectors can be saved as an image. The difference of a raster and a vector is shown in Figure \ref{fig:raster2vector}. A raster is a grid of pixels, where each pixel has its colour value. During the conversion, the shape will be formulated as a vector information hence produce better graphics. The raster pixels' colour information is used to identify the edge of the area, which is then processed into a vector, for example, a line or a circle. This process will cause loss of details due to resolution issue, whereas the output is satisfying the Geofence requirement. 

\begin{figure}[H]
	\includegraphics[scale=1]{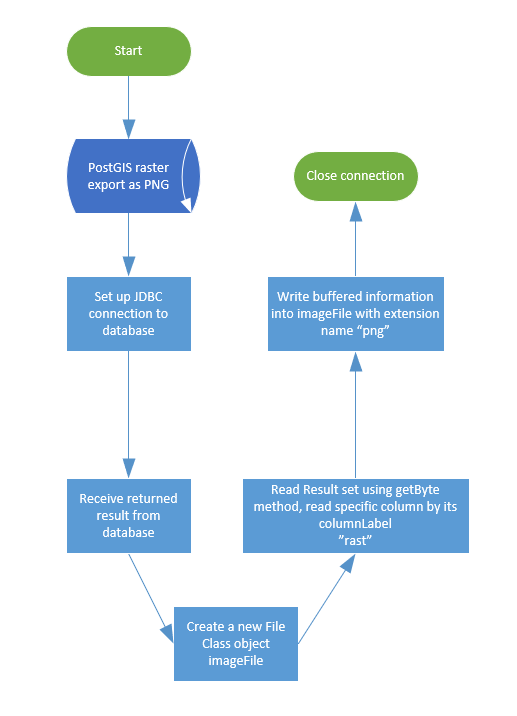}
	\caption{Flowchart exporting png file from PostGIS raster }
	\label{fig:png}
\end{figure}

The export process is done using PostGIS built-in PNG/TIFF/JPG export function, then the JDBC access the database and retrieve it as byte information. The Java "File" class write the byte information into a PNG file to finish the exportation. The work flow of above operation is shown in Figure \ref{fig:png}

\subsubsection{Geofence Advisory external Interface}
\begin{figure}[H]
	\centering
	\includegraphics[scale=0.9]{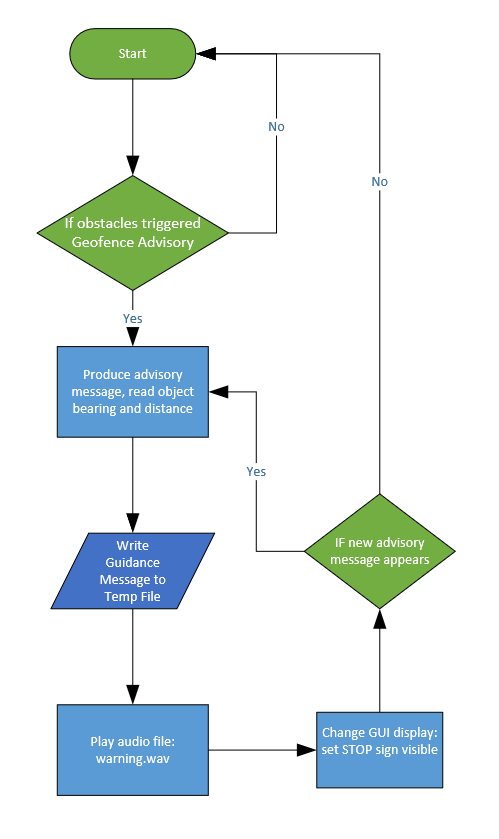}
	\caption{Flow Chart: Geofence Advisory to external resource}
	\label{fig:adflow}
\end{figure}

The advisory produce the guidance information for operator further consideration. The Audio alert is played immediately after the event is triggered. Meanwhile, a text based message will be save to source location for the controller. The flow chart in Figure \ref{fig:adflow} shows the temp file is created when the advisory is produced.

\section{Geofence Function Development(Geofence Classes and Methods)\label{sec:class}}
In this section, all the major functions of Geofence Software are described using flowchart with detailed explanation. The source code for each function is supplied as a snippet of code, which only reflects its own work. The complete code is supplied in the source file submission.

In the following text, functions start with ST is a PostGIS function, and be used exclusively in PostGIS.
\subsection{Geofence Reference Map Display}
\begin{figure}[H]
	\includegraphics[scale=0.8]{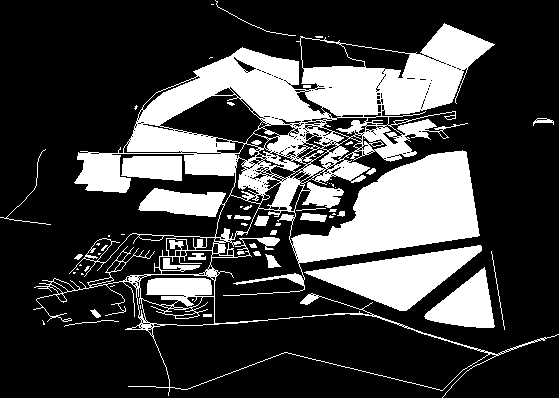}
	\caption{The reference Map display around Cranfield Campus}
\end{figure}
\begin{figure}[H]
	\includegraphics[scale=0.8]{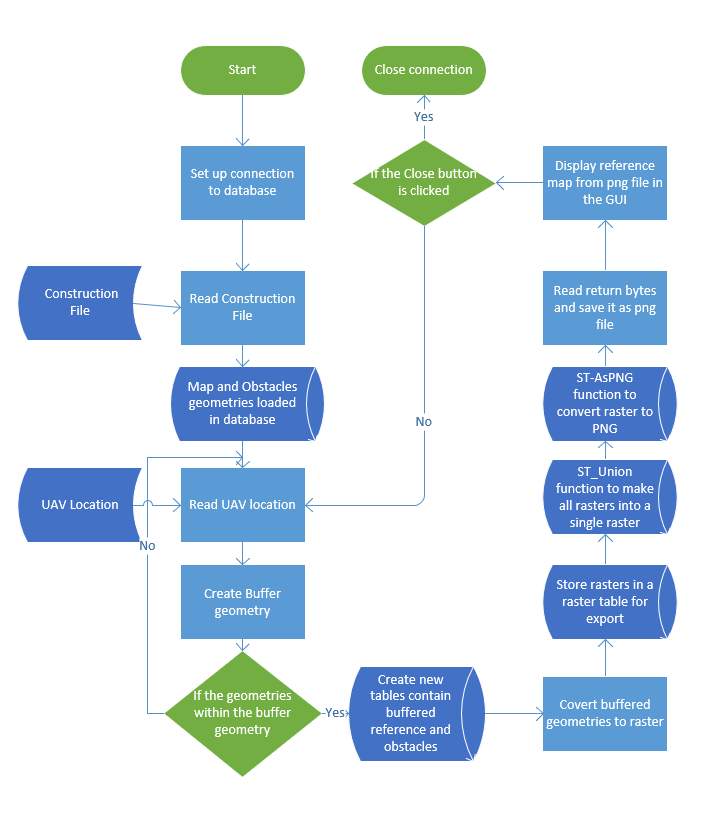}
	\caption{Flowchart: Produce Geofence reference map display}
\end{figure}

\subsection{Geofence Obstacle layer Display}
The obstacle image is rendered from the obstacle raster, which is produced by the obstacles' geometry. The product from raster conversion in reference map also includes the obstacles, but stored in a different table. When the ST\underline{  }Union function puts all obstacle rasters into a single raster, the different colour is applied. The colour code for obstacle is red, in the form of RGB array [255,0,0]. The obstacle raster references the location and size from the reference map, and hence becomes an overlay for the reference map.

\begin{figure}[H]
	\includegraphics[scale=1]{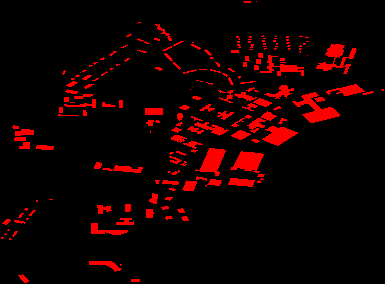}
	\caption{Obstacle overlay}
	\label{fig:obstacle}
\end{figure}

The obstacle display reads the obstacle image from the file system and shows it on the GUI, as Figure \ref{fig:obstacle}.

\begin{figure}[H]
	\centering
	\includegraphics[scale=0.5]{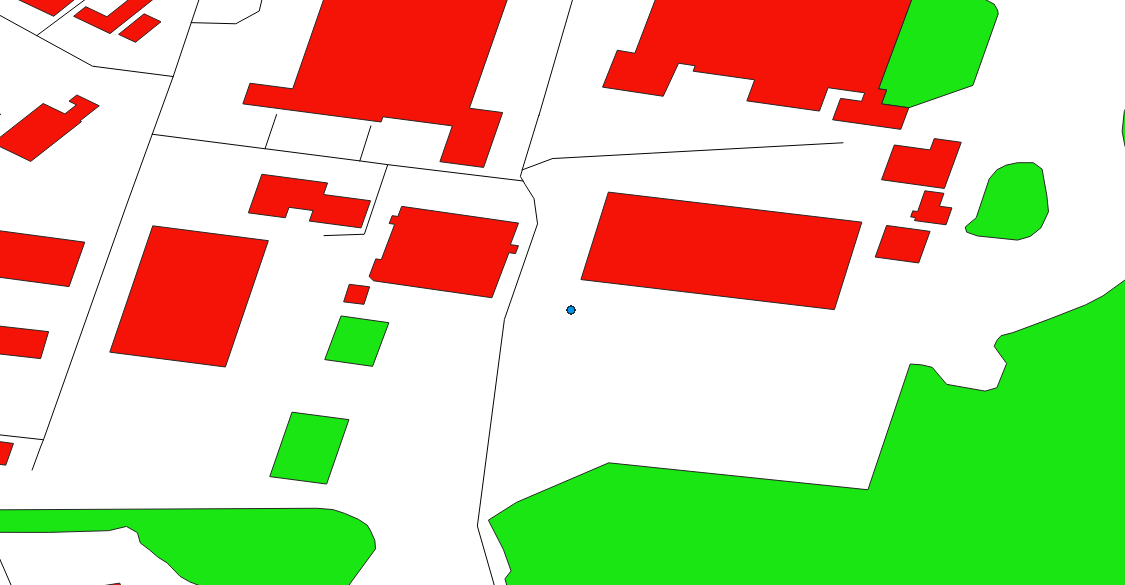}
	\caption{A layer of Obstacles and Reference Map with transparent background}
	\label{fig:stnthetic}
\end{figure}

The display function combines the image layers from reference and the obstacles. The synthetic view shown in Figure \ref{fig:stnthetic}.

\begin{figure}[H]
	\centering
	\includegraphics[scale=0.81]{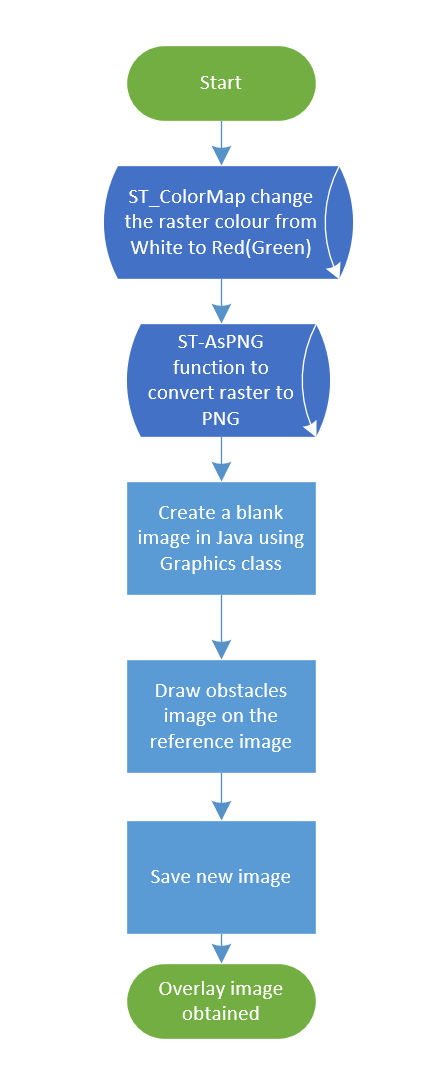}
	\caption{Flowchart: Making an overlay of reference and obstacles}
	\label{fig:overlay}
\end{figure}

The process of making a sytnthetic view from two rasters is described in Figure \ref{fig:overlay}. In this flowchart, the obstacle raster was produced in previous process, hence only the colour adjustfiacation is required. The same logic is also used to render and display the open area, which is assigned to green colour. 
\subsection{Geofence Advisory Audio Alert and Sign Alert}
\begin{figure}[H]
	\centering
	\includegraphics[scale=1]{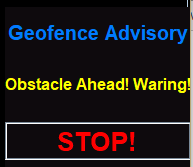}
	\caption{Visual alert from the advisory function}
	\label{fig:alert}
\end{figure}

While the situation is displayed, the advisory works in the background. Advisory function provides the UAV operator with audio and visual alert. An example of Geofence alert is shown in Figure \ref{fig:alert}. In this example, the yellow font indicates the emerging threat from current situation, and the red "STOP" font means the immediate action is required, meanwhile, the stop control message is generated for flight controller.

\begin{figure}[H]
	\centering
	\includegraphics[scale=1]{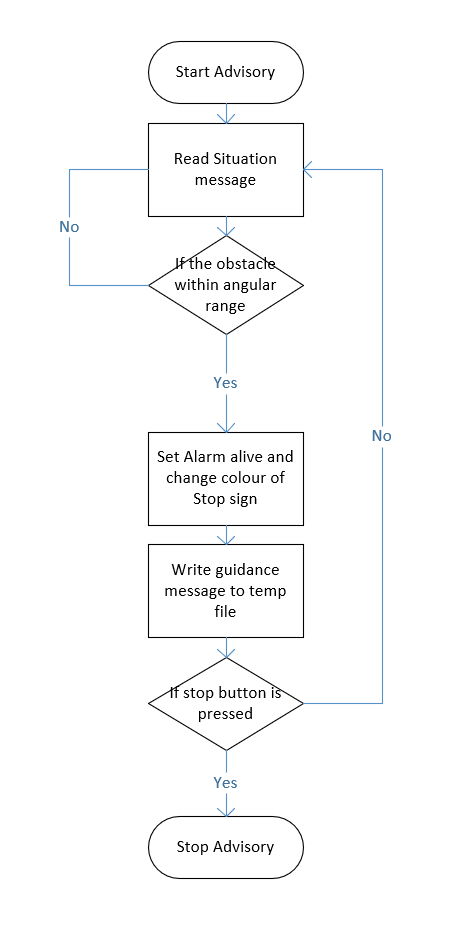}
	\caption{Flowchart: The decision making for advisory alerts}
	\label{fig:decision}
\end{figure}

The program will decide whether the advisory is triggered using the algorithms demonstrated in Appendix \ref{sec:code-for-geo-fence-advisory-message-generation}. The flowchart showing the decision-making process is in Figure \ref{fig:decision}

The advisory function uses UAV heading as the judgement criteria. Geofence Advisory sets a virtual cone for the UAV and points the bottom to the flying direction, as it is explained in Figure \ref{fig:ad3}. Any object within the cone will trigger the alarm and change the colour of STOP sign from invisible to red. The algorithm to find if the object is within the angular range is also described in Appendix \ref{sec:code-for-geo-fence-advisory-message-generation}, where a multi branch choice is used.

\subsection{Geofence Input and Output}
The Geofence has two input sources, the construction file and the UAV location source. The construction file is design to be a plain text, comma-separated values. The Geofence program reads the text, parse values into a array. Then the constructor function will deliver these values into the database for further operation.

UAV location is also supplied as a string type value, besides reading the value from the text, the simulator also provides the input interface for users inputting simulated values.

\section{Database Design}
\begin{figure}[H]
	\centering
	\includegraphics[scale=0.61,angle=90]{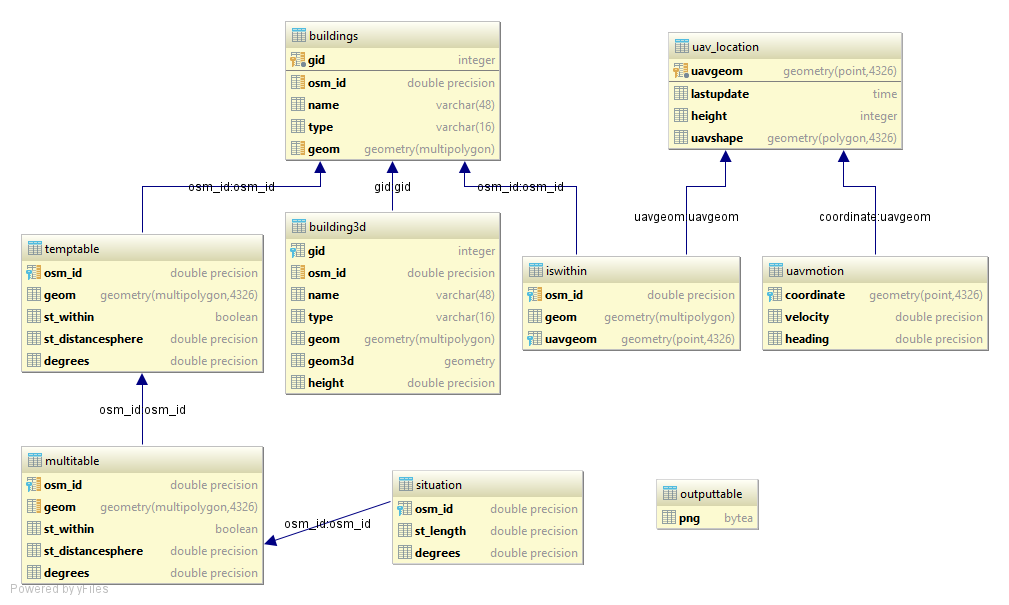}
	\caption{Example Database Diagram of the Geofence}
	\label{fig:database}
\end{figure}

There are nine tables in the Geofence Database. Following description is the design principle, which does not 100\% match the real production environment. Nine tables are illustrated in Figure \ref{fig:database}, where table: buildings uav\underline{  }location and uavmotion, have raw input from the map providers and UAV GNSS and sensors. 

Table "building3d" is an auxiliary table for the 3-dimensional Geofence. It uses the value from another external source of information, Height table, to generate a 3d box. However, in this prototype system, the 3d object is not the main target due to lack of data. It serves as the reserved table for future improvement.

Table "iswithin" stores the UAV geometry and the buildings geometries from "buildings" table, it caches all the geometries data for the further process.

Table "temptable" is designed to keep all the boolean values and geometries from the "iswithin" table and the result of "st\underline{  }within" function. It provides cached data for the further results set.

Table "multitable" is the result set of all items in the buffer zone. It also supplies the data to the table "situation" to help reduce the process time of searching a greater table.

Table "outputtable" stores the raster and png data from the "multitable", the reason why it is not linked by the foreign key is that the raster or png is the result of a union operation. The union operation combines all the geometries in "multitable" and output a raster or png for the output purpose. 

\begin{figure}[H]
	\centering
	\includegraphics[scale=0.6]{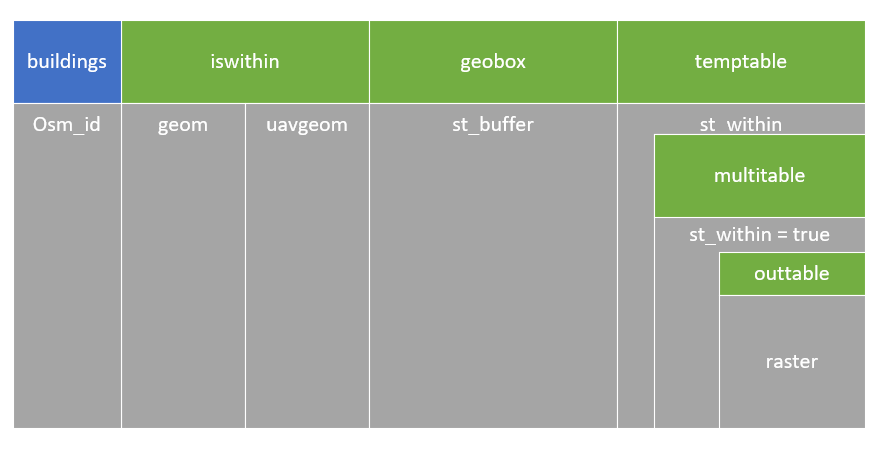}
	\caption{Table relation of the Geofence Database}
	\label{fig:relation}
\end{figure}

An example table relation of using buildings as obstacles is shown in Figure \ref{fig:relation}. Bule and green blocks mean the table name, and the grey area indicates the column name. The geom column, which is not shown in the figure, is shared by all tables, in order to let the geom reach the outtable.

The detail creation commands of all tables can be found in Appendix \ref{appendix6}
\section{Compilation}
To compile the Geofence GUI and rest of the classes, the Java complier is used to generate a .jar File, which is an executable package file. A JAR file allows the Java Runtime to run the whole program in a well manner and automatically includes used classes and associated files. The build information is stored in the META-INF/MANIFEST.MF file, which is supplied in the source submission.

\chapter{Geo-fence Software Testing and Discussion\label{cp:6test}}
The testing of the Geofence is divided into two parts, the functional testing and the performance testing. In the functional testing session, the configuration of the software will be tested using a simulated UAV location near Cranfield University Cranfield Campus. The certain tasks will be performed:
\begin{itemize}
	\item Set Geofence according to UAV location
	\item Configure Geofence to include more types of objects into geofence
	\item Display Geofenced area with the desired buffer area
	\item Show divert alert
	\item Save distance and bearing to the geofence
	\item Provide simulation based on simulated UAV location
	
\end{itemize}
\section{Geo-fence software hardware set up}

\begin{figure}[H]
	\centering
	\includegraphics[scale=0.9]{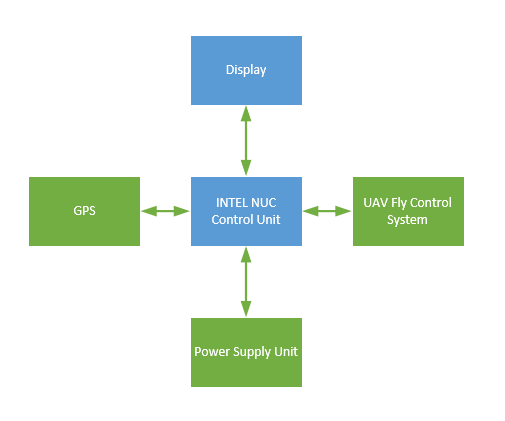}
	\caption{Hardware Composition of the rig}
\end{figure}

The testing platform is a NUC by Intel, detail specification:\newline
CPU: I7 5557U @3.1GHz\newline
Memory : 16 GB DDR3 1600MHz\newline
Drive : PCIe SSD \newline

\begin{figure}[H]
	\centering
	\includegraphics[scale=0.7]{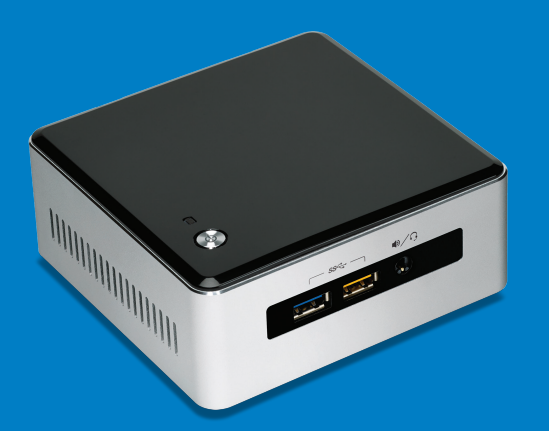}
	\caption{An Intel NUC\cite{Intel}}
	\label{fig:nuc}
\end{figure}

\section{Geo-fence software configuration}
\subsection{Required Software}
The Geofence software has its dependencies, which include the runtime and the database management system with spatial data extension.\newline
The following explicit software are required for compilation.\newline
Database : PostgreSQL v10 \newline
Database extension : PostGIS 2.4\newline
Java SDK : Java SE 10 (jdk-10.0.2)\newline

The Java SDK is not required to run the Geofence, hence a separate list for the Geofence User who only runs the software.\newline
Database: PostgreSQL v10, SQLite V3.24.0\newline
Database Extension: PostGIS 2.4\newline
Java Runtime: Java SE 10(jre-10.0.2)\newline

The above software is required to run the geofence, however, a lower or higher version of the runtime is not tested, and may result in a lack of functions and may cause errors.
\subsection{Initial Set-up}
The Geofence database is part of the initial preparation, it has to be loaded before the Geofence software is used. The example of making a area into database is shown below in Figure \ref{fig:postgresql}

\begin{figure}[H]
	\centering
	\includegraphics[scale=0.7]{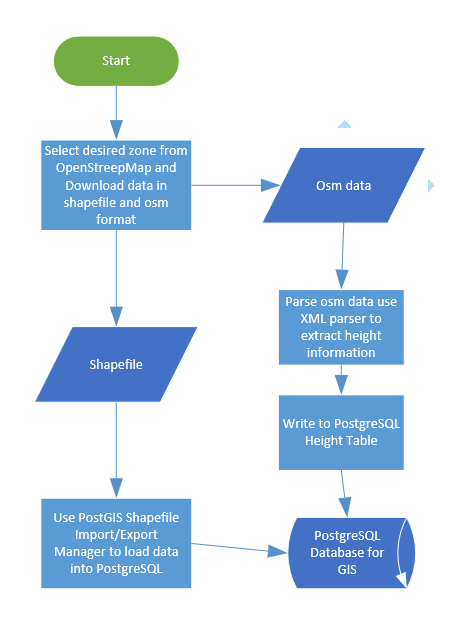}
	\caption{Example of making a postgreSQL database for GIS data}
	\label{fig:postgresql}
	
\end{figure}

Figure \ref{fig:postgresql} shows a way to transfer GIS data, however, many methods can achieve the same goal. A image of database is provided in the project source file, as the part of the submission. The user should use postgreSQL v10 to load the database for best result replication.

An mirror image of used database is supplied together to reduce the work to verify the design. Additional, the simulator has the initialisation function to help user build the database from scratch.

\subsection{Simulation Scenarios}
To simulate the UAV flying around Cranfield University campus, following parameters are selected to reflect a possible situation: \newline
Test 1:\newline
Heading: 355 degree towards North\newline
Horizontal Velocity : 8 m / s\newline
Flying height : 30 m\newline
Latitude : 52.073 (with omissions)\newline
Longitude : -0.627 (with omissions)\newline

\begin{figure}[H]
	\includegraphics[scale=0.75]{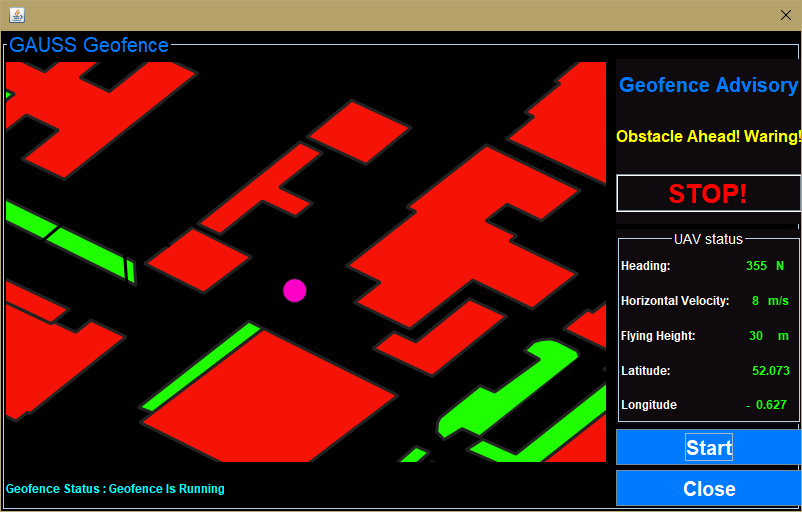}
	\caption{A working snapshot of UAV display from simulation}
	\label{fig:uavgui}
\end{figure}

Geofence is set to mark all buildings as obstacles, and showing them as red geometry shapes.\newline
The UAV is represented as Purple dot.
A working Geofence simulation on UAV hand-held device shall look like Figure \ref{fig:uavgui}, which displays the result from first test.

Test 2:\newline
Heading: 355 degree towards North\newline
Horizontal Velocity : 10 m / s\newline
Flying height : 30 m\newline
Latitude : 52.80 (with omissions)\newline
Longitude : -0.625 (with omissions)\newline

\begin{figure}[H]
	\includegraphics[scale=0.7]{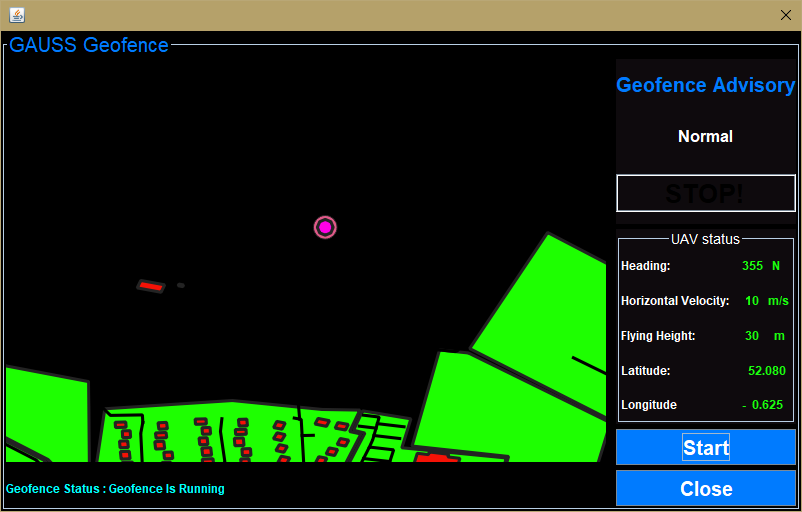}
	\caption{Test 2 Geofence Display}
	\label{fig:test2}
\end{figure}

The result for the Geofence Situation and Geofence Advisory is shown in Figure \ref{fig:test2}. In the second test, the north space from the UAV location is open. The UAV can fly towards north without triggering Geofence Advisory. 

\section{Performance Analysis}
The performance of the system consists of the hardware specification and algorithm. To analysis the performance on the different device, another PC is introduced, detail specification is listed below :\newline
CPU: I7 4770 @3.4GHz\newline
Memory : 16 GB DDR3 1600MHz\newline
Drive : Kingston SHFS37A240G SSD \newline

The test procedure is to simulate the Geofence Software normal uses. In this test case, a 200 second running time is executed and the resource usage and response time are recorded by JProfiler, a Java Virtual Machine Monitoring Tool.

The test result is exclusive to this test rig and a second test, which based on the Intel NUC is also exclusive to the Intel NUC only.

\begin{figure}[H]
	\includegraphics[scale=1]{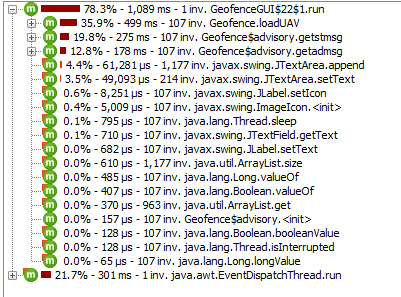}
	\caption{The CPU time allocation of Geofence Software}
	\label{fig:test1}
\end{figure}

In Figure \ref{fig:test1}, the bar length indicates how long the class and methods called the CPU. In GeofenceGUI class, the main logic of the geofence is performed, loadUAV method took the most of time to be called, due to the reason that a raster image was produced every time it was called. 

\begin{figure}[H]
	\includegraphics[scale=1]{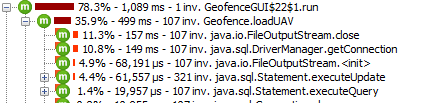}
	\caption{Content inside loadUAV method execution}
	\label{fig:loadUAV}
\end{figure}

Inside loadUAV method, in Figure \ref{fig:loadUAV}, the FileIO and connection to the database is the major consumption of the resource.

\begin{figure}[H]
	
	\includegraphics[scale=1]{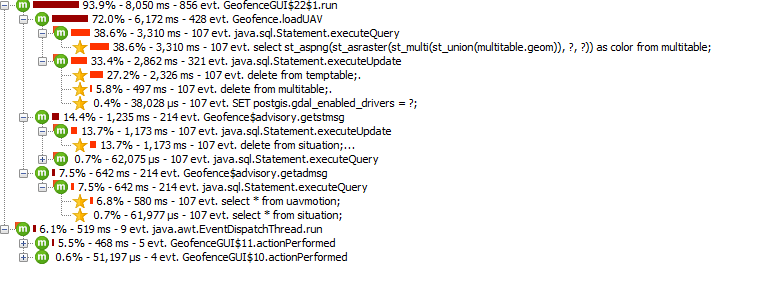}
	\caption{The CPU time percentage allocation of database related operation}
	\label{fig:testdatabase}
\end{figure}

\begin{figure}[H]
	\includegraphics[scale=1]{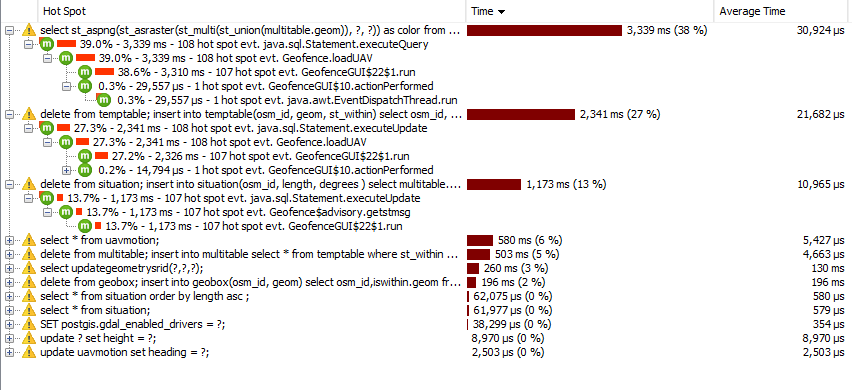}
	\caption{The average execution time of database related operation}
	\label{fig:testdatabasetime}
\end{figure}

Hence the main performance related operation is the data I/O, the closer look of the database connection is demonstrated in Figure \ref{fig:testdatabase}
From Figure \ref{fig:testdatabasetime}, the most time consuming operation is the raster production. In this query, the PostGIS union two tables and make a raster from them. The elements in each tables affect the work load of the operation, and it leads to the test of different buffer size.

The buffer size is the critical element that affects the performance, the different buffer sizes are used in the test. The result of 5 sets of configurations are shown in Table \ref{tab:buffer}. The time shown in the Table \ref{tab:buffer} is the execution time of SQL command:

\begin{lstlisting}[language = sql]
delete from temptable;
insert into temptable(osm_id, geom, st_within) 
select osm_id, geom,st_within(geom, st_buffer) 
as 
st_within from geobox,uav_location;"
\end{lstlisting}

\begin{table}[H]
	\centering
	\caption{Time consumption with different buffer size selected}
	\begin{tabular}{|c|c|}
		\hline 
		Buffer Size in degree & Execution Time in ms \\ 
		\hline 
		0.05 & 2,240 \\ 
		\hline 
		0.02 &  717\\ 
		\hline 
		0.01 &  215\\ 
		\hline
		0.005 & 77\\
		\hline
		0.002 & 30\\
		\hline 
	\end{tabular} 
\label{tab:buffer}
\end{table}
\begin{figure}
	\centering
	\includegraphics[scale=1]{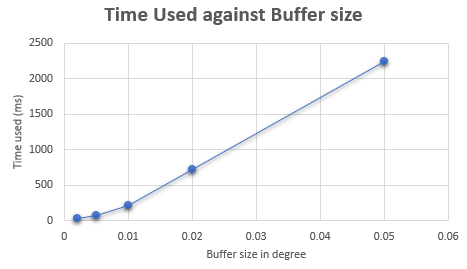}
	\caption{Time Used against Buffer size}
	\label{fig:bvt}
\end{figure}
In Figure \ref{fig:bvt}, a liner relation is established between the time consumption and buffer size. It proves the buffer size has major impact on performance. Moreover, a larger buffer size means more objects in the buffered zone, which is dependant on area of selection.

To conclude, the buffer size affects the performance of the software most, where the area of selection may result different response time, but the trend of increased time with larger buffer zone is clear.

A second test with above settings will be conducted on the Intel NUC platform where the result may differ due to hardware specification difference. The result is included in Appendix \ref{appendix:test}. The Intel NUC has a lot faster respond time thanks to its ultra-fast m.2 type SSD storage. The similar result certifies the relation between performance and buffer size. 

It is worth notice that the fast storage can minimise the execution time and allow larger buffer zone to be loaded. Meanwhile, the refresh rate can be improved to sub-second level to maximise the security.

\section{Comparison with similar service}
The products with similar purpose and functions to this project are Google Geofencing API and Apple Region Monitoring API of Core Location. The comparison of them is shown in table \ref{tab:com} 
\begin{table}[H]
	\caption{The comparison chart of similar service providers}
\begin{tabular}{p{2.8cm}|p{2cm}|p{2.5cm}|p{2cm}|p{2cm}|p{2cm}}
	
	\hline 
	&By Coordinates &By Attributes of Object& World Wide & Off-Line Operation & Cross-Platform \\ 
	\hline 
	This Project & \checkmark & \checkmark & \checkmark *conditions & \checkmark & \checkmark \\ 
	\hline 
	Google Geofencing API & \checkmark & $\times$ & \checkmark & $\times$ & \checkmark \\ 
	\hline 
	Region Monitoring API & \checkmark & \checkmark & \checkmark & $\times$ & $\times$ \\ 
	\hline 
	Flight Planner & \checkmark & $\times$ & \checkmark & \checkmark& \checkmark\\
	\hline
	DJI Controller & \checkmark & $\times$ & $\times$ & $\times$& \checkmark
	\label{tab:com}
\end{tabular} 
\end{table}
*condition: A worldwide service depends on the map providers and local authorities. Buildings height is supplied separately. 

Besides, this project offers unique functionalities comparing with commercial API mentioned above. It provides UAV-related guidance message to the user and offers the surrounding situation awareness.

\subsection{Advantages}

In the public developer documentation, there is no comparable service, for example, setting up geofence using existing elements on the map. 

The unique feature of the GAUSS Geofence Software is the power to build the geofence with operators demand. The fence can be defined by building's height, land use, building type and many more attributes available in the OpenStreetMap or OS Map data.  Geofence is generated automatically using pre-defined script.

\subsection{Disadvantages}
Service provider, such as Google and Apple, have their commercial coverage of whole world. The map data is richer in details, such as building heights, building type or even business information. They also have stable service to all the users, whereas this project is only using an unreliable static map server from OpenStreetMap.de. The response speed is a lot faster while the Google Map API is used.

\chapter{Conclusions and Future Updates\label{cp:7con}}
\section{Conclusions}
UAV safety and ground objects safety are secured by employing Geofence Software system. With Geofence software installed, the UAV retains in the safe zone whenever there is connection loss or signal spoofing.

The functionality of the Geofence software satisfies the requirement of this project. It has achieved geofencing entry and exit management (Geofence), geofence collision avoidance (Advisory) and geofence-guidance-navigation display fusion (Navigation). 

 The breakthrough of setting Geofence by the object's attributes such as building type, landuse, owner, etc. The customised geofence construction is exclusive to this software and it is a valuable and an important feature of this Geofence Software Project.

The Geofence software has achieved all objectives with lots of opportunity to improve it. The architecture of the Geofence software is flexible and extensible.

\section{Limitations}
Limitations on the software functionality and development concern the successful implementation of the project.

From the development process, a thoughtful robust code is always demanded, whereas the limited time results not enough work on the code quality. 

Though all the input values are checked, the null pointer issue of Java is not checked thoroughly. 

\subsection{Height Integration}
The height in the map model has been simplified with only buildings and objects' height, regardless of the terrain elevation and ref system error. The drawback of such can be minimised by restricting the projected usage inside a certain region or disable the height parameter in the Geofence. Besides, the public released terrain elevation only offers 3 arc-second, which is around 300 meter for each tile. Such terrain elevation is a potential threat to the precision of the geo-fence. It should not be used unless the 1 arc second tiles are available.

In another word, the limitation is not technical but lack of accurate GIS resource.

\subsection{Software Development Process}
The project is driven by the functions it carries, the development process is lacking a systematic software engineering planning process, as it is the nature of Aerospace Vehicle Design Course. The product needs to be improved by software development conventions,i.e. Software Engineering, in order to comply with the software life cycle. 

Debugging is not part of the Geofence project, therefore the quality of the code needs to be verified against the industrial standard.

\subsection{Separation of PostGIS Server and SQLite Server }
The initial plan of making the database system for the Geofence software is to put GIS data in the PostGIS and the vector file in the SQLite database due to its integration level design. This will help the hand-held terminal use less resource and also make rendering and visualisation more dedicated from the main program. It will allow better extensible architecture to be used.
\section{Future Updates}
\subsection{Dynamic Buffer Size}
The dynamic buffer size is considered to have better performance due to the fact that the smaller buffer is, the faster the fence is. In future, a buffer zone should change according to its speed, for example, the slow flying UAV with 20m per second velocity needs much less buffered zone than fast flying one. A 20 second window results 400m for the slower one, but 2000m for fast one at 100m/s.

\subsection{3D object integration}
This project aimed to implement 3D objects at the beginning, however, only partially fulfilled. The reason of having a 3D object instead of a 2D object with additional height data is to enable more spatial estimations and operations between UAVs and the fenced objects. For example, the distance from a UAV is measured as the perpendicular distance from the UAV coordinate to the polygon shape in 2D set up, and absolute distance in 3D. Also, the possibility to use the Geofence inside a building or a fenced object completely relies on the 3D fencing technology, where the fence inside the fence could occur.

The database is designed to contain 3D objects, hence, once the accurate data is ready, the integration should be smooth and easy.

\begin{figure}[H]
	\includegraphics[scale=0.7]{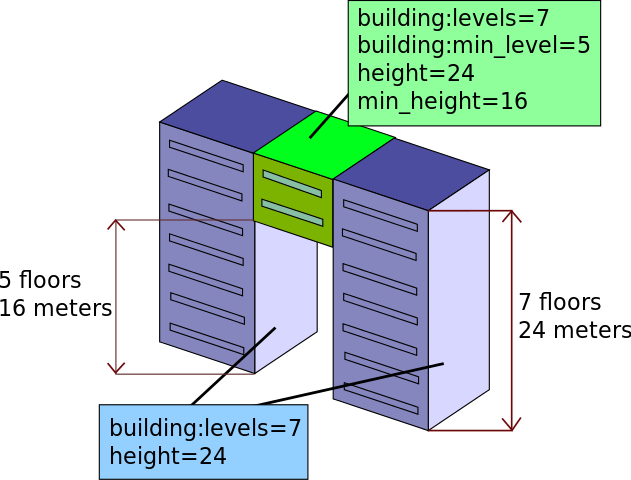}
	\caption{A example of 3D building construction\cite{OpenStreetMapContributors2017}}
	\label{fig:building}
\end{figure}

A typical 3D building, shown in Figure \ref{fig:building}, is the working target in the future updates to allow the UAV flow across the open part of the building, which is also similar to the bridge tunnel.

\subsection{Geofence generation using script}
The pre-defined settings can be included in a script file and read by the Geofence Software to provide a bespoke geofence solution. The effort of making a script is very little, but the purpose of making it is slightly off the scope. As a further development, a more vanilla process of using the Geofence Software, a script parser and maker is recommended. 

\subsection{Geofence Software Manual}
Due to lack of the time, a manual for Geofence Software is not present. The future work on making the manual is encouraged.

\subsection{Geofence Advisory export JSON/XML file}
In this stage of Geofence Software, the interfacing file with external controller is limited to strings, stored as a plain text file. In the further stage, a XML or JSON object as the interface object is recommended. It will benefit the rest of device who needs to read or write from and to the Geofence Software.
\subsection{Airspace Integration}
The geofence software not only provides the flat geofencing service but also covers the elevation cases. The airspace information, managed by the authority, can be a great source of 3D geofencing. The height attribute can be used to create a column with given coordinates. These require a bit of working on airspace integration, on how the interface works between each other. GAUSS Geofence Software System is ready to accept it.

\bibliography{irp}
\bibliographystyle{plain}

\appendix
\chapter{Test result of Intel NUC hardware platform\label{appendix:test}}
Intel NUC test result of executing following SQL command using a series of buffer size:
\begin{lstlisting}[language = sql]
delete from temptable;
insert into temptable(osm_id, geom, st_within) 
select osm_id, geom,st_within(geom, st_buffer) 
as 
st_within from geobox,uav_location;"
\end{lstlisting}
\begin{table}[H]
	\centering
	\caption{Time consumption with different buffer size selected on Intel NUC}
	\begin{tabular}{|c|c|}
		\hline 
		Buffer Size in degree & Execution Time in ms \\ 
		\hline 
		0.05 & 46 \\ 
		\hline 
		0.02 &  29\\ 
		\hline 
		0.01 &  19\\ 
		\hline
		0.005 & 17\\
		\hline
		0.002 & 15\\
		\hline 
	\end{tabular} 
	\label{tab:nuc}
\end{table}

\chapter{Code for Database Generator}
\begin{lstlisting}[language = Java, caption = Geofence Database Builder]

import java.sql.*;

public class DatabaseBuilder {
String url = "jdbc:postgresql://127.0.0.1/postgres";
String usr = "postgres";
String pwd = "123";
Statement stmt = null;
Connection c = null;

public void LoadSql() {
try {
c = DriverManager.getConnection(url, Geofence.usr, pwd);
System.out.println("Connection Success");
stmt = c.createStatement();
ResultSet rt = stmt.executeQuery("select * from information_schema.tables where tables.table_schema ='public';");
while (rt.next()) {
System.out.println(rt.getString(3));
}


stmt.close();
c.close();
} catch (Exception e) {
System.err.println(e.getClass().getName() + ":" + e.getMessage());
System.out.println("Database error");
System.exit(0);
}
}

public void loadGeofence() {
Connection c = null;
System.out.println("Start Reading Database...");
//Here should be a connection to Postgresql Database

try {
c = DriverManager.getConnection(url, Geofence.usr, pwd);
System.out.println("Database connected");
stmt = c.createStatement();
System.out.println("Creating Geofence tables");
int rt = stmt.executeUpdate(
"create table IF NOT EXISTS uav_location (\n" +
"  lastupdate time,\n" +
"  height     int\n" +
");\n" +
"create table IF NOT EXISTS uavmotion (velocity float,heading float,height float);" +
"-- iswithin table store the value for st_within function to compare.\n" +
"create table IF NOT EXISTS situation\n" +
"(\n" +
"  osm_id  double precision,\n" +
"  length  double precision,\n" +
"  degrees double precision\n" +
");");
try {
ResultSet rs1=stmt.executeQuery("select addgeometrycolumn('uavmotion','coordinate',4326,'point',2);");
ResultSet rs2=stmt.executeQuery("select addgeometrycolumn('uav_location', 'uavgeom', 4326, 'point', 2);");
ResultSet rs3=stmt.executeQuery("select addgeometrycolumn('uav_location', 'uavshape', 4326, 'polygon', 2);");
System.out.println("column added");
} catch (SQLException e) {
e.printStackTrace();
System.out.println("column ok");
}
int rt1 = stmt.executeUpdate(
"update uavmotion set coordinate = (select uavgeom from uav_location);" +
"create table IF NOT EXISTS iswithin as" +
"  select buildings.osm_id, buildings.geom, uav_location.uavgeom\n" +
"  from buildings,\n" +
"       uav_location;"
+ "create table IF NOT EXISTS temptable\n"
+ "(\n"
+ "  osm_id    double precision,\n"
+ "  geom      geometry(MultiPolygon, 4326),\n"
+ "  st_within boolean);"

+ "create table IF NOT EXISTS multitable\n"
+ "(\n"
+ "  osm_id            double precision,\n"
+ "  geom              geometry(MultiPolygon, 4326),\n"
+ "  st_within         boolean\n"
+ ");"
+ "create table IF NOT EXISTS uavmotion\n"
+ "(\n"
+ "  velocity   double precision,\n"
+ "  heading    double precision,\n"
+ "  coordinate geometry(Point, 4326),"
+ "height       double precision"
+ ");");
ResultSet rs4 = stmt.executeQuery("select updategeometrysrid('geobox', 'geom', 4326);");
ResultSet rs5 = stmt.executeQuery("select updategeometrysrid('buildings', 'geom', 4326);\n");
ResultSet rs6 = stmt.executeQuery("select updategeometrysrid('landuse', 'geom', 4326);\n");
ResultSet rs7 = stmt.executeQuery("select updategeometrysrid('roads', 'geom', 4326);\n");
ResultSet rs8 = stmt.executeQuery("select updategeometrysrid('waterways', 'geom', 4326);\n");
ResultSet rs9 = stmt.executeQuery("select updategeometrysrid('railways', 'geom', 4326);\n");
System.out.println("Success");
stmt.close();
c.close();
} catch (Exception e) {
System.err.println(e.getClass().getName() + ":" + e.getMessage());
System.out.println("Database error");
}
}

public void GeofenceConstructer() throws SQLException {
Connection d = null;
System.out.println("Start Reading Database...");
//Here should be a connection to Postgresql Database
String url = "jdbc:postgresql://127.0.0.1/postgres";
String usr = "postgres";
String pwd = "123";
try {
d = DriverManager.getConnection(url, usr, pwd);
System.out.println("Connection Success");
} catch (Exception e) {
System.err.println(e.getClass().getName() + ":" + e.getMessage());
System.out.println("Database error");
System.exit(0);
}

try {
Statement st = d.createStatement();
int rt = st.executeUpdate("" +
"CREATE EXTENSION postgis; " +
"CREATE EXTENSION postgis_topology;" +
" CREATE EXTENSION postgis_sfcgal;" +
"  CREATE EXTENSION fuzzystrmatch;" +
"  CREATE EXTENSION address_standardizer;" +
" CREATE EXTENSION address_standardizer_data_us;" +
" CREATE EXTENSION postgis_tiger_geocoder;" +
" CREATE EXTENSION pgrouting;" +
" CREATE EXTENSION ogr_fdw;" +
"  CREATE EXTENSION pointcloud;" +
" CREATE EXTENSION pointcloud_postgis;");
System.out.println("Database with postgis created");
} catch (Exception e) {
System.err.println(e.getClass().getName() + ":" + e.getMessage());
System.out.println("Database already created");
}
}
}

\end{lstlisting}
\chapter{Code for Geo-fence Advisory Message Generation}\label{sec:code-for-geo-fence-advisory-message-generation}
\begin{lstlisting}[language=Java, caption=Java Geofence Advisory Message Generation]
 public static class advisory {
 
 public ArrayList getstmsg() {
 Connection c = null;
 stmt = null;
 ArrayList infoList = new ArrayList();
 try {
 c = DriverManager.getConnection(url, usr, pwd);
 System.out.println("Connection Success");
 stmt = c.createStatement();
 int rt = stmt.executeUpdate("delete from situation;\n" +
 "insert into situation(osm_id, length, degrees ) select bufferedobstacle.osm_id,\n" +
 "       st_length(st_shortestline(st_transform(st_setsrid(bufferedobstacle.geom, 4326), 27700),\n" +
 "                                 st_transform(st_setsrid(uav_location.uavgeom, 4326), 27700) ))as length,\n" +
 "       degrees(st_azimuth(st_centroid(bufferedobstacle.geom), uav_location.uavgeom))\n" +
 "from uav_location,\n" +
 "     bufferedobstacle;");
 ResultSet rs = stmt.executeQuery("select * from situation order by length asc ;");
 while (rs.next()) {
 ResultSetMetaData rsm = rs.getMetaData();
 int id = rs.getInt("osm_id");
 double degree = rs.getDouble("degrees");
 String degreeString = String.format("%.0f", degree);
 double length = rs.getDouble("length");
 String lengthString = String.format("%.2f", length);
 String infoString = ("Object OSM ID: " + id + " at degree:" + degreeString + " with distance of " + (lengthString) + " meter" + System.lineSeparator());
 System.out.println(infoString);
 
 infoList.add(infoString);
 }
 stmt.close();
 c.close();
 
 
 } catch (Exception e) {
 System.err.println(e.getClass().getName() + ":" + e.getMessage());
 System.out.println("Database error");
 System.exit(0);
 }
 return infoList;
 }
 
 public ArrayList getadmsg() {
 Connection c = null;
 stmt = null;
 Statement stmt1 = null;
 ArrayList adList = new ArrayList();
 String adString = new String();
 try {
 c = DriverManager.getConnection(url, usr, pwd);
 System.out.println("Connection Success");
 System.out.println("Reading headings");
 stmt = c.createStatement();
 stmt1 = c.createStatement();
 
 ResultSet rs1 = stmt.executeQuery("select * from uavmotion;");
 ResultSet rs2 = stmt1.executeQuery("select * from situation;");
 
 while (rs1.next()) {
 Double heading = rs1.getDouble("heading");
 System.out.println("heading read success");
 while (rs2.next()) {
 Double degree = rs2.getDouble("degrees");
 if (heading <= 350) {
 if ((heading - 10) < degree && degree < (heading + 10)) {
 System.out.println("Make diversion");
 String degreeString = String.format("%.0f", degree);
 adString = "Make diversion to aviod going " + degreeString + " degree \n";
 adList.add(adString);
 }
 } else if (heading < 10 && heading >= 0) {
 if (degree > (350 + heading) || degree < (10 + heading)) {
 System.out.println("Make diversion");
 String degreeString = String.format("%.0f", degree);
 adString = "Make diversion to aviod going " + degreeString + " degree \n";
 adList.add(adString);
 
 }
 } else if (heading > 350 && heading <= 360) {
 if ((heading - 10) < (degree + 360) && (degree + 360) < (heading + 10)) {
 System.out.println("Make diversion");
 String degreeString = String.format("%.0f", degree);
 adString = "Make diversion to aviod going " + degreeString + " degree \n";
 adList.add(adString);
 }
 }
 }
 }
 stmt.close();
 c.close();
 } catch (Exception e) {
 e.printStackTrace();
 System.err.println(e.getClass().getName() + ":" + e.getMessage());
 System.out.println("Database error");
 System.exit(0);
 }
 return adList;
 }
 }
 
\end{lstlisting}

\chapter{PostgreSQL table creation command\label{appendix6}}
\begin{lstlisting}[language = sql]
-- enable gdal drivers for png output
set postgis.gdal_enabled_drivers = 'enable_all';
-- buildings table exists before the program starts, it is supplied by the map provider

alter table buildings
add unique (osm_id);
select st_setsrid(buildings.geom, 4326)
from buildings;
-- add table building3d for 3d fencing
select * into building3d
from buildings;
alter table building3d
add unique (osm_id);
alter table "building3d"
add constraint "fk_building3d_gid_osm_id_name_type_geom" foreign key ("gid")
references "buildings" ("gid");
alter table building3d
add geom3d geometry;
alter table building3d
add height float;
update building3d
set geom3d = st_force3d(geom);
select st_setsrid(geom3d, 4326);
-------------------------------------
-- add table height for 3d fencing reference.
select gid, osm_id into "height"
from buildings;
alter table "height"
add height float;
alter table "height"
add constraint "fk_height_gid_osm_id" foreign key ("osm_id")
references "building3d" ("osm_id");
-------------------------------------
-- add table of uav location
create table uav_location (
lastupdate time,
height     int
);
-- uavgeom is the point location of uav;
-- uavshape is the geometry of the size of a uav
select addgeometrycolumn('uav_location', 'uavgeom', 4326, 'point', 2);
select addgeometrycolumn('uav_location', 'uavshape', 4326, 'polygon', 2);
-- a testing value for uav_location
insert into uav_location (lastupdate, height, uavgeom, uavshape)
values (current_time,
30,
st_setsrid(st_makepoint(-0.627365, 52.072989), 4326),
st_buffer(st_setsrid(st_makepoint(-0.627365, 52.072989), 4326),
0.00001, 'quad_segs=2'));

-- the uavgeom is unique and is used as the primary key here
alter table uav_location
add constraint "pk_uav_location" primary key ("uavgeom");

-- uavmotion table contains the uav heading and velocity for geofence advisory
select uav_location.uavgeom as coordinate into uavmotion
from uav_location;
alter table uavmotion
add velocity float;
alter table uavmotion
add heading float;
alter table uavmotion
add constraint "pk_uavmotion" foreign key ("coordinate") references uav_location (uavgeom);

-------------------------------------
-- iswithin table store the value for st_within function to compare.
create table iswithin as
select buildings.osm_id, buildings.geom, uav_location.uavgeom
from buildings,
uav_location;


delete
from iswithin;
insert into iswithin (osm_id, geom)
SELECT osm_id, geom
from buildings;
update iswithin
set uavgeom = uav_location.uavgeom
from uav_location;

alter table iswithin
add unique (osm_id);
alter table "iswithin"
add constraint "fk_iswithin_osm_id_geom" foreign key ("osm_id")
references "buildings" ("osm_id");
alter table "iswithin"
add constraint "fk_iswithin_uavgeom" foreign key ("uavgeom")
references "uav_location" ("uavgeom");

drop table iswithin cascade;
-------------------------------------
-- geobox is alternative name for buffered zone, where the centre location is the uav and the radius is defined by the degree in wgs84 system

create table geobox as
select iswithin.osm_id, st_buffer((select iswithin.uavgeom from uav_location), 0.012), iswithin.geom
from iswithin;
select updategeometrysrid('iswithin', 'geom', 4326);
select updategeometrysrid('geobox', 'geom', 4326);
select updategeometrysrid('buildings', 'geom', 4326);
delete
from geobox;
insert into geobox (osm_id, geom)
select osm_id, iswithin.geom
from iswithin;
update geobox
set st_buffer = st_buffer((select uav_location.uavgeom
from uav_location), 0.002);


alter table geobox
add unique (osm_id);

-- st_area to return the size of the buffered zone
select st_area(st_transform(st_buffer, 27700))
from geobox;
alter table "geobox"
add constraint "fk_geobox_osmid" foreign key ("osm_id")
references "iswithin" ("osm_id");


drop table geobox cascade;
-------------------------------------
-- temptable stores the boolean value from st_within function and surronding objects' bearing and distance

select osm_id, geom, st_within(geom, st_buffer)
into temptable
from geobox;
--
delete
from temptable;
insert into temptable (osm_id, geom, st_within)
select osm_id, geom, st_within(geom, st_buffer)
from geobox;

-- select st_srid(geom) from geobox;
-- select st_srid(geom) from iswithin;
-- select st_srid(geom) from buildings;
-- select st_srid(uavgeom) from iswithin;
-- select st_srid(st_buffer) from geobox;
-- select st_within(geom, st_buffer) from geobox,uav_location;
-- select updategeometrysrid('geobox','geom',4326);
--
--
-- update temptable set osm_id=geobox.osm_id,
--        geom=geobox.geom,
--        st_within=st_within(geobox.geom, geobox.st_buffer)
-- from geobox,
--      uav_location;


alter table temptable
add unique (osm_id);
alter table temptable
add unique (geom);
alter table "temptable"
add constraint "fk_temptable_osm_id_geom" foreign key ("osm_id")
references "buildings" ("osm_id");

-------------------------------------
-- multitable gathers all the geomtries within the radius of the buffer zone geobox and prepare them to be unioned into a multipolygon
select * into multitable
from temptable
where st_within = true;

delete
from multitable;
insert into multitable
select *
from temptable
where st_within = true;

alter table multitable
add unique (geom);
alter table multitable
add unique (osm_id);
alter table "multitable"
add constraint "fk_multitable_osm_id_geom" foreign key ("osm_id")
references "temptable" ("osm_id");
-------------------------------------
-- outputtable stores all the ready to display image from the multitable results
set postgis.gdal_enabled_drivers = 'enable_all';
select st_aspng(st_asraster(st_multi(st_union(multitable.geom)), 500, 500)) as png into outputtable
from multitable;

alter table "outputtable"
add constraint "fk_outputtable_asraster" foreign key ("png")
references "multitable" ("geom");
-------------------------------------
-- situation table contains the bearing from uav to building object and the distance from uav to buildings
select multitable.osm_id,
st_length(st_shortestline(st_transform(st_setsrid(multitable.geom, 4326), 27700),
st_transform(st_setsrid(uav_location.uavgeom, 4326), 27700)))as length,
degrees(st_azimuth(st_centroid(multitable.geom), uav_location.uavgeom)) into situation
from uav_location,
multitable;

alter table "situation"
add constraint "fk_situation_geom" foreign key ("osm_id")
references "multitable" ("osm_id");

delete
from situation;
insert into situation (osm_id, length, degrees)
select multitable.osm_id,
st_length(st_shortestline(st_transform(st_setsrid(multitable.geom, 4326), 27700),
st_transform(st_setsrid(uav_location.uavgeom, 4326), 27700)))as length,
degrees(st_azimuth(st_centroid(multitable.geom), uav_location.uavgeom))
from uav_location,
multitable;

\end{lstlisting}

\end{document}